\documentclass[reprint,aip,jcp,graphicx]{revtex4-1}

\usepackage{amssymb}
\expandafter\let\csname equation*\endcsname\relax

\expandafter\let\csname endequation*\endcsname\relax

\usepackage{mathtools,amsmath}
\usepackage{graphicx}
\usepackage{marginnote}
\usepackage{color}
\usepackage{siunitx}
\usepackage{changes}
\usepackage{xcolor}
\usepackage{hyperref}
\hypersetup{
	colorlinks   = true,
	citecolor    = darkblue,
	linkcolor    = darkblue,
	urlcolor     = darkblue
}
\usepackage{bbm}
\usepackage{times,txfonts}
\usepackage{esdiff}

\definecolor{markedgreen}{HTML}{000000}
\definecolor{darkblue}{rgb}{0.,0.,0.5}
\definecolor{darkred}{rgb}{0.7,0.,0.}

\DeclareMathAlphabet{\pazocal}{OMS}{zplm}{m}{n}

\begin{document}

\title{\Large Pushing the limits of the reaction-coordinate mapping
\newline}

\author{Luis A. Correa}
\affiliation{CEMPS, Physics and Astronomy, University of Exeter, Exeter, EX4 4QL, United Kingdom}
\affiliation{School of Mathematical Sciences and CQNE, University of Nottingham, University Park Campus, Nottingham NG7 2RD, United Kingdom}
\affiliation{Kavli Institute for Theoretical Physics, University of California, Santa Barbara, CA 93106}

\author{Buqing Xu}
\affiliation{School of Mathematical Sciences and CQNE, University of Nottingham, University Park Campus, Nottingham NG7 2RD, United Kingdom}

\author{Benjamin Morris}
\affiliation{School of Mathematical Sciences and CQNE, University of Nottingham, University Park Campus, Nottingham NG7 2RD, United Kingdom}

\author{Gerardo Adesso}
\affiliation{School of Mathematical Sciences and CQNE, University of Nottingham, University Park Campus, Nottingham NG7 2RD, United Kingdom}

\date{\today}

\begin{abstract}
	The reaction-coordinate mapping is a useful technique to study complex quantum dissipative dynamics into structured environments. In essence, it aims to mimic the original problem by means of an `augmented system', which includes a suitably chosen collective environmental coordinate---the `reaction coordinate'. This composite then couples to a simpler `residual reservoir' with short-lived correlations. If, in addition, the residual coupling is \textit{weak}, a simple quantum master equation can be rigorously applied to the augmented system, and the solution of the original problem just follows from tracing out the reaction coordinate. \textit{But, what if the residual dissipation is strong?} Here we consider an exactly solvable model for heat transport---a two-node linear ``quantum wire'' connecting two baths at different temperatures. We allow for a structured spectral density at the interface with one of the reservoirs and perform the reaction-coordinate mapping, writing a perturbative master equation for the augmented system. We find that: (a) strikingly, the stationary state of the original problem can be reproduced accurately by a weak-coupling treatment even when the residual dissipation on the augmented system is very strong; (b) the agreement holds throughout the entire dynamics under large residual dissipation in the overdamped regime; (c) and that such master equation can grossly overestimate the stationary heat current across the wire, even when its non-equilibrium steady state is captured faithfully. These observations can be crucial when using the reaction-coordinate mapping to study the largely unexplored strong-coupling regime in quantum thermodynamics.  
\end{abstract}

\maketitle

\section{Introduction}\label{sec:intro}

Understanding the dynamics of open quantum systems in structured environments is central to nearly all aspects of quantum research---from modelling the chemistry of biomolecules \cite{garg1985effect,hartmann2000controlling,roden2012accounting}, to understanding the thermodynamics of quantum systems \cite{binder2018thermodynamics}, or assisting in the design of nano-structures for quantum-technological applications \cite{alicki2004optimal,mccutcheon2010quantum,higgins2013thermometry}. Unfortunately, treating open systems in complex environments is extremely challenging, the main reason being the absence of a clear-cut timescale separation between system and evironmental dynamics \cite{devega2017nonmarkovian}. Various tools exist to deal with such problems, including exact path-integral methods \cite{feynman2000theory,hu1992brownian,tanimura1990heom}, stochastic Schr\"{o}dinger equations \cite{stockburger2002nonmarkov,devega2005multiple}, unitary transformations \cite{wagner1986unitary,wurger1998polaron}, or Markovian embeddings \cite{garraway1997pseudomodes,martinazzo2011universal,woods2014mappings,iles2014environmental}. Here, we shall focus on the latter; specifically, on the ``reaction-coordinate mapping'' \cite{nazir2018RCmapping}.

In a seminal paper by Garg \textit{et al.} \cite{garg1985effect} a very simple \textit{ansatz} was put forward for the structure of the environment modulating the rate of an electron-transfer process in a biomolecule. Essentially, it assumes that a distinct collective environmental coordinate---the \textit{reaction} coordinate (RC)---couples strongly to the donor--acceptor system, which can be thought-of as a two-level spin. In this construction, the combined effect of all other environmental degrees of freedom would merely cause semiclassical \textit{friction} on the spin--RC composite. It is then possible to view the spin as an open system and work out its dissipative dynamics via, e.g., exact path-integral methods.

Interestingly, the ansatz can be ``turned on its head'' \cite{hughes2009effective,martinazzo2011universal,iles2014environmental} and viewed as a Markovian embedding technique. Namely, an arbitrarily complicated environment may be iteratively decomposed by first, extracting a collective environmental coordinate and working out the coupling of the resulting `augmented system' to the remaining `residual environment'. By repeating this procedure sufficiently many times, one ends up with an open-system model with the simplest friction-like Ohmic dissipation \cite{martinazzo2011universal,woods2014mappings}, albeit with a much larger system size. Whenever the residual friction (i.e., dissipation strength) is perturbatively small, the problem can be rigorously solved via standard weak-coupling Markovian master equations \footnote{A Fokker--Plank equation may be derived in the opposite large-friction limit \cite{garg1985effect,iles2014environmental}.}. This provides a simple route to tackle otherwise intractable open quantum systems, especially when a single iteration of the procedure suffices for the problem at hand.

The reaction-coordinate mapping has been applied extensively to open quantum systems strongly coupled to both bosonic \cite{iles2014environmental,iles2016energy,strasberg2016nonequilibrium,restrepo2018quantum,wertnik2018optimizing,maguire2018environmental,puebla2019spin,martensen2019transmission,mcconnell2019electron,lambert2019virtual} and fermionic \cite{strasberg2018fermionic,schaller2018electronic,restrepo2019electron} reservoirs. Its relative ease of use and the neat physical picture that emerges from it, in terms of, e.g., system--environment correlation-sharing structure \cite{iles2014environmental,iles2016energy}, make it particularly appealing as a general-purpose open-system tool. Unfortunately, relying on perturbative master equations imposes \textit{a priori} severe limitations on the parameter ranges in which the method can be used. Intriguingly, however, it has resisted benchmarking at finite temperatures over a wide friction range \cite{iles2014environmental,iles2016energy,strasberg2018fermionic}, which made us wonder where are its true limitations \footnote{Recently, the RC mapping has been shown to break down at large friction in the zero-temperature limit \cite{lambert2019virtual}. Our calculations here are, however, limited to \textit{finite} temperatures.}.

In this paper, we set out precisely to ``push'' the method to the limit, by deliberately taking the forbidden large friction limit in a minimal heat-transport setup. Our biggest advantage is that we work with an exactly solvable model \cite{gonzalez2017testing}; we can thus always benchmark the accuracy of the mapping without having to approximate the exact dynamics numerically. Under \textit{steady-state} conditions, we find that the RC mapping does work accurately even under extremely large friction, in spite of the fact that the underlying master equation breaks down. We also find that overdamped \textit{dynamics}, resulting in strong residual friction, is accurately captured by this method. Importantly, however, when the residual friction is strong and one relies on weak-coupling master equations to compute heat (or particle) currents across the non-equilibrium open system of interest, the results can be completely flawed and yet, appear physically consistent. This observation can have important consequences when using the reaction-coordinate mapping to explore the thermodynamics of strongly coupled nanoscale open systems; \textit{verifying that the method approximates the state of an open system correctly is certainly not enough to trust it with the calculation of quantum-thermodynamic variables}.

As a by-product of our master-equation analysis of the augmented system subjected to friction, we derive here a (global) Born--Markov secular quantum master equation for a general linear network of harmonic nodes coupled to arbitrarily many equilibrium environments. This generalises the customarily used \textit{local} master equations applied to quantum transport problems through weakly interacting networks \cite{asadian2013heat}. We also write the ensuing non-equilibrim steady state, and explicit formulas for the corresponding stationary heat currents. Finally, we discuss the \textit{dos and don'ts} of the often confusing Hamiltonian frequency-renormalisation counter-terms that appear in quantum Brownian motion \cite{caldeira1983path,ford1988qle,weiss2008quantum}, as it is particularly important to use them consistently when performing the reaction-coordinate mapping. 

This paper is structured as follows: In Sec.~\ref{sec:quantum_wire}, we introduce our simple model and discuss very briefly the reaction-coordinate mapping. In Sec.~\ref{sec:GME}, we provide the general quantum master equation that we shall later apply on our augmented system. Rather than reproducing the standard textbook derivation from the microscopic system--bath(s) model, we limit ourselves to provide here the key steps, and write down instead the full equations of motion explicitly, along with their stationary solutions, and the corresponding steady-state heat currents. In Sec.~\ref{sec:QLE} we outline the exact solution of both our original problem and that of the augmented system undergoing (arbitrarily strong) friction. We then proceed to discuss the steady-state (cf. Sec.~\ref{sec:steady-state-and-heat-currents}) and dynamical (cf. Sec.~\ref{sec:dynamics}) benchmarks to the reaction coordinate mapping, commenting both on the approximation to the \textit{state} of the system and to the \textit{stationary heat currents} flowing across it. Finally, in Sec.~\ref{sec:conclusions}, we wrap up and draw our conclusions.

\section{The model and its solution}\label{sec:model-and-solution}

\subsection{A two-node non-equilibrium quantum wire}\label{sec:quantum_wire}

\begin{figure*}[t!]
	\centering
	\includegraphics[width=0.55\linewidth]{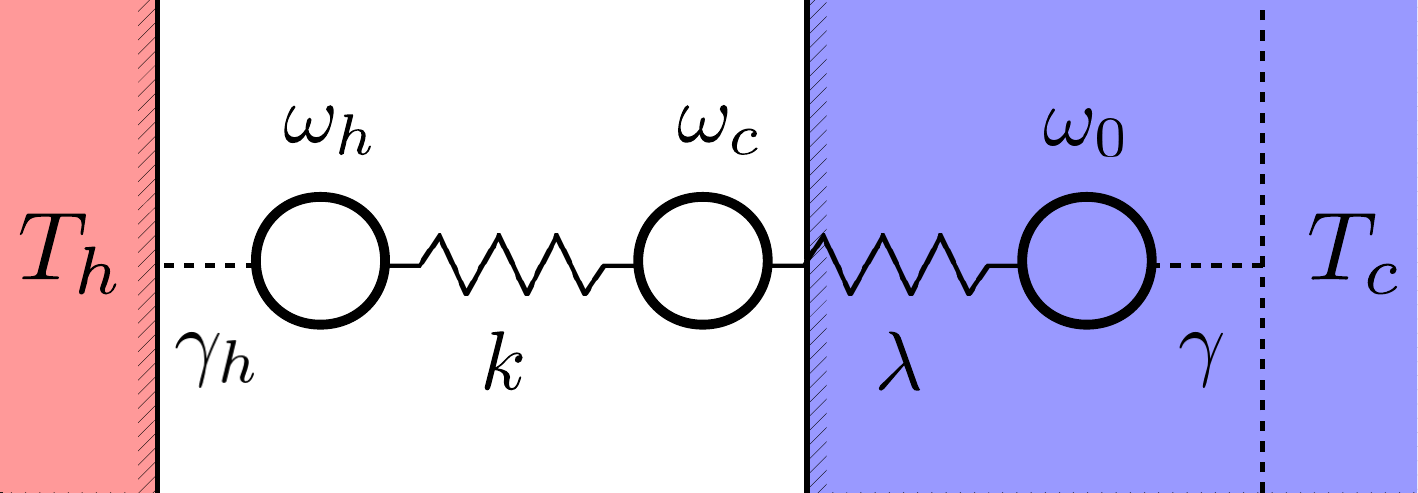}
	\caption{Sketch of the non-equilibrium quantum wire with nodes at frequencies $\omega_h$ and $\omega_c$ and internal coupling $k$. The dissipative interaction between node $\omega_h$ and the corresponding (hot) bath, at $ T_h $, is characterized by an Ohmic spectral density, e.g., $ J_h(\omega) \sim \gamma_h\, \omega $. As a result, the corresponding environmental correlation time is short. Furthermore, the dissipation strength $ \gamma_h $ is assumed to be perturbatively weak. On the contrary, the (cold) bath at $T_c$ features long-lived correlations due to the structured spectral density $J_c(\omega)=\gamma\lambda^2\omega/\big[\gamma^2\omega^2+\big(\omega^2-\omega_0^2\big)^2\big]$. The resulting dynamics can be mimicked exactly by coupling a reaction coordinate at frequency $\omega_0$ to the system with strength $\lambda$. This composite makes up the augmented system which, in turn, couples to a residual reservoir---customarily assumed to be in equilibrium also at $T_c$---via $ \tilde{J}_c(\omega) \sim \gamma\,\omega $. This guarantees that the residual environmental correlations for the augmented system are short-lived. However, if the `friction' coefficient $\gamma$ in $J_c(\omega)$ is large, so is the residual dissipation strength. Crucially, this clashes with the weak-coupling approximation which underpins \textit{any} perturbative quantum master equation that could be written for the augmented three-node system.} \label{fig:0}
\end{figure*}

\subsubsection{Full Hamiltonian}\label{sec:full_hamiltonian}

As already advanced, our model consists of a two-node chain (or ``quantum wire'') of harmonic oscillators with a linear spring-like coupling of strength $ k $ (see Fig.~\ref{fig:0}), that is
\begin{equation}
\pmb{H}_w = \sum\nolimits_{\alpha\,\in\,\{h,c\}}\left(\frac12\omega_\alpha^2 \pmb{X}_\alpha^2 + \frac{\pmb P^2_\alpha}{2}\right) + \frac{k}{2}(\pmb{X}_h-\pmb{X}_c)^2.
\label{eq:hamiltonian_wire}
\end{equation}
Note that here and in what follows, we set all masses to one. We shall also take $ \hbar = k_B = 1 $. The wire is kept out of equilibrium by two linear bosonic baths at temperatures $ T_\alpha $. Throughout, $ \alpha \in \{h,c\} $ stands for `hot' or `cold', i.e., $ T_h > T_c $. Their Hamiltonians can thus be cast as $ \pmb H_{T_\alpha} = \sum\nolimits_{\mu} \omega_\mu\, \pmb{a}_\mu^{(\alpha)\,\dagger}\pmb{a}^{(\alpha)}_\mu $, where $ \pmb{a}^{(\alpha)\,\dagger}_\mu $ ($ \pmb{a}_\mu^{(\alpha)} $) is a creation (annihilation) operator of bath $ \alpha $ in the collective bosonic environmental mode at frequency $ \omega_\mu $. In turn, the dissipative interactions between the wire and the baths are 
\begin{equation}
\pmb{H}_{\text{diss},\,\alpha} = \pmb X_\alpha\otimes \pmb B_\alpha \coloneqq \pmb{X}_\alpha \sum\nolimits_{\mu} g_\mu^{(\alpha)}\,\pmb x^{(\alpha)}_\mu \qquad \alpha\in\{h,c\},
\label{eq:hamiltonian_wire-bath}
\end{equation}
where the quadratures $ \sqrt{2 \omega_\mu}\,\pmb x^{(\alpha)}_\mu \coloneqq \pmb a_\mu^{(\alpha)\,\dagger} + \pmb a_\mu^{(\alpha)} $ and, as usual, the coupling constants $ g_\mu^{(\alpha)} $ make up the spectral densities
\begin{equation}
J_\alpha(\omega) \coloneqq \,\pi \sum\nolimits_{\mu} \frac{g_\mu^{(\alpha)\,2}}{2 \omega_\mu}\delta(\omega-\omega_\mu) \qquad \alpha\in\{h,c\}\,.
\label{eq:spectral_density_general}
\end{equation}

Importantly, each system--bath coupling $ \pmb H_{\text{diss},\alpha} $ requires us to introduce a renormalisation term in the bare Hamiltonian of the wire $ \pmb H_w \mapsto \pmb H_w + \pmb\delta\pmb H_{w\text{--}\alpha} $, which compensates for the environmental distortion on the system's potential \cite{weiss2008quantum}. If we were not to include such terms and let $ T_h = T_c = T $ be arbitrarily large, the exact stationary state would approach $ \pmb\varrho_w(\infty)\sim\exp{[-(\pmb H_w - \pmb\delta\pmb H_{w\text{--}h} - \pmb\delta\pmb H_{w\text{--}c})/T ]} $ instead of the classical limit $ \pmb\varrho_w(\infty)\sim\exp{(-\pmb H_w/T)} $; this should be seen as an important deficiency of the model \cite{caldeira1983path}. Specifically, these extra terms are
\begin{equation}
\pmb\delta \pmb H_{w\text{--}\alpha} = \pmb X_\alpha^2\,\sum\nolimits_{\mu} \frac{g_\mu^{(\alpha)\,2}}{2\omega_\mu^2} = \pmb X_\alpha^2\,\int_0^\infty\frac{\text{d}\omega}{\pi}\,\frac{J_\alpha(\omega)}{\omega} \coloneqq \frac{\delta_\alpha}{2}\pmb X_\alpha^2,
\label{eq:renormalisation}
\end{equation}
and the full Hamiltonian of our system is, therefore,
\begin{equation}
\pmb H = \pmb H_{T_h} + \pmb{H}_{\text{diss},\,h} + \pmb\delta\pmb H_{w\text{--}h} + \pmb H_w + \pmb\delta\pmb H_{w\text{--}c} + \pmb H_{\text{diss},\,c} + \pmb H_{T_c}.
\label{eq:original_hamiltonian}
\end{equation}

We take an Ohmic spectrum for the coupling to the `hot bath', i.e., $ J_h(\omega) = \gamma_h\,\omega\,\theta(\omega/\Lambda_h) $, where $ \theta(x) $ is some rapidly decaying function for arguments $ x > 1 $, which places an upper bound on the excitation energies. For practical reasons we choose the algebraic cutoff $ \theta(x) = (1+x^2)^{-1} $, although other choices would not alter our results as long as $\Lambda_h$ is large. Such $ J_h(\omega) $ is referred-to as `overdamped' in the context of energy transfer in molecular systems \cite{iles2016energy}. For the coupling of the wire to the cold bath, we take instead the `underdamped' spectrum
\begin{equation}
J_c(\omega) = \frac{\gamma\,\lambda^2\,\omega}{\gamma^2\omega^2+(\omega^2-\omega_0^2)^2},
\label{eq:spectral_density_structured}
\end{equation}
which displays a peak around $ \omega_0 $, whose height and width are essentially controlled by $ \lambda $ and $ \gamma $, respectively \footnote{Note that in the limit of very large $ \gamma $ this becomes $ J_c(\omega) \sim \frac{\lambda^2}{\gamma\,\omega} $.}. This is precisely the effective spectral density resulting from the aforementioned \textit{ansantz} by Garg \textit{et al.} \cite{garg1985effect}. The frequency-renormalisation shifts $ \delta_\alpha $ for these expectral densities are explicitly given by $ \delta_h = \gamma_h\Lambda_h $ and $ \delta_c = \lambda^2/\omega_0^2 $.

The decay of the environmental correlation functions $ \left\langle \pmb B_\alpha(t)\,\pmb B_\alpha(0) \right\rangle $ gives an idea of the bath's memory time, and to which extent a simple Markovian relaxation process can be a good approximation to the actual dynamics. Specifically \cite{Breuer2002}
\begin{equation}
\left\langle \pmb B_\alpha(t)\,\pmb B_\alpha(0) \right\rangle = \int_0^\infty \frac{\text{d}\omega}{\pi}~J_\alpha(\omega)\left( \coth{\frac{\omega}{2T_\alpha}}\cos{\omega t} - i \sin{\omega t} \right).
\label{eq:environmental-correlation}
\end{equation}
While{\color{markedgreen}, at finite temperatures,} a spectral density like our $ J_h(\omega) $ typically leads to very short correlation times, consistent with the Markovian approximation, a spectrum such as \eqref{eq:spectral_density_structured} can give rise to very long-lived correlations and thus, to a much more complex dynamics. {\color{markedgreen} However, at sufficiently low temperatures---a regime which we shall not explore here---the bath correlation times can become comparable to the typical system dynamics even for an Ohmic spectral density.}

\subsubsection{The reaction-coordinate mapping in a nutshell} \label{sec:rc-mapping-nutshell}

To circumvent this problem one may try to exploit the fact that Eq.~\eqref{eq:spectral_density_structured} is the effective spectral density for a system which couples \textit{indirectly}---namely, through a bosonic mode, or reaction coordinate, of frequency $ \omega_0 $---to a residual reservoir with a purely Ohmic spectrum \cite{garg1985effect}; the coupling between the auxiliary mode and the system being of strength $ \lambda $ (see Fig.~\ref{fig:0}). Put in other words, the dynamics 
\begin{equation}
\frac{\mathrm{d}}{\mathrm{d} t}\, \pmb{\varrho}_w(t) = -i\,\text{tr}_{\bar{w}}\,[\tilde{\pmb H},\tilde{\pmb{\rho}}]
\label{eq:dynamics_augmented}
\end{equation}
generated by
\begin{multline}
\tilde{\pmb H} \coloneqq  \pmb H_{T_h} + \pmb H_{\text{diss},\,h} + \pmb\delta\pmb H_{w\text{--}h} + \pmb H_w + \pmb\delta\pmb H_{w\text{--}c} \\ 
- \lambda\,\pmb{X}_c\pmb X_\text{RC} + \frac12\left(\omega_0^2\,\pmb X_\text{RC}^2 + \pmb{P}_\text{RC}^2\right) + \pmb\delta\pmb H_{\text{RC--res}} \\
 + \pmb X_\text{RC} \sum\nolimits_{\mu} \tilde{g}_\nu^{(c)} \tilde{\pmb{x}}^{(c)}_\nu + \sum\nolimits_\nu \omega_\nu\,\tilde{\pmb a}_\nu^{(c)\,\dagger}\tilde{\pmb a}_\nu^{(c)} ,
\label{eq:full_Hamiltonian}
\end{multline}
exactly coincides with that of
\begin{equation}
\frac{\text{d}}{\text{d}t}\,\pmb \varrho_w(t) = -i\,\text{tr}_{\bar{w}} [\pmb{H}, \pmb{\rho}],
\label{eq:dynamics_original}
\end{equation}
when the coefficients $ \{ g^{(c)}_\mu \} $ in $ \pmb H_{\text{diss},\,c} $ correspond to Eq.~\eqref{eq:spectral_density_structured} [by virtue of \eqref{eq:spectral_density_general}] and the $ \{\tilde{g}_\nu^{(c)}\} $ in the sixth term on the right-hand side of Eq.~\eqref{eq:full_Hamiltonian}, to $ \tilde{J}_c(\omega) = \gamma\,\omega $; technically, some suitable cutoff function $ \theta{(\omega/\tilde{\Lambda}_c)} $ with would be required, the mapping being exact only in the limit $ \tilde{\Lambda}_c \rightarrow \infty $. Here, $ \text{tr}_{\bar{w}} $ amounts to tracing over all degrees of freedom except for the wire. The boldface symbols with tilde correspond to operators completely or partly supported in the residual reservoir; in our case, the quadratures $ \{ \tilde{\pmb x}_\nu^{(c)} \} $, the creation and annihilation $ \{ \tilde{\pmb a}_\nu^{\dagger\,(c)},\tilde{\pmb a}_\nu^{(c)} \} $ operators in modes at frequency $ \omega_\nu $; and the joint state of the hot bath, the wire, the reaction coordinate, and the residual reservoir $ \tilde{\pmb\rho}(t) $. Finally, the newly introduced operators $ \pmb X_\text{RC} $ and $ \pmb P_\text{RC} $ stand for the canonical degrees of freedom of the RC. Note that we have included as well the renormalisation term $ \pmb\delta\pmb H_\text{RC--res} $ arising from the coupling between the RC and the residual bath [cf. Eq.~\eqref{eq:renormalisation}]. Accessible and rigorous derivations of the equivalence between Eqs.~\eqref{eq:dynamics_original} and \eqref{eq:dynamics_augmented} can be readily found in the literature \cite{garg1985effect,iles2014environmental,martinazzo2011universal,strasberg2016nonequilibrium}.

There is, however, an important caveat regarding the \textit{initial condition} for the augmented system. It is common practice to assume that the residual reservoir is in equilibrium at temperature $ T_c $, just like the original physical bath (see Fig.~\ref{fig:0}); and to initialise the auxiliary RC in a thermal state at $ T_c $, uncorrelated from the rest \cite{iles2014environmental,iles2016energy}. Note that the dynamics generated by Eqs.~\eqref{eq:dynamics_augmented} and \eqref{eq:dynamics_original} only agree if $ \pmb\rho(0) = \tilde{\pmb\rho}(0) $, i.e., 
\begin{equation}
\pmb{\varrho}_{T_h}\,\otimes\,\pmb\varrho_w(0)\,\otimes\,\pmb\varrho_{T_c} = \pmb\varrho_{T_h}\,\otimes\,\pmb\varrho_w(0)\,\otimes\,\tilde{\pmb\varrho}_{\text{RC + res}}(0)\,.
\end{equation}
In particular, this means that the composite `RC + residual reservoir' should start instead in a \textit{joint thermal state} at temperature $ T_c $; that is, $ \tilde{\pmb{\varrho}}_\text{RC + res}(0) = \pmb\varrho_{T_c} $, which is \textit{not} of the form $ \pmb{\varrho}_\text{RC}(0)\otimes\tilde{\pmb\varrho}_{T_c} $. Hence, there could be large initial correlations between the RC and the residual reservoir, especially at low $ T_c $. Importantly, the absence of correlations with the environment is central to the derivation of the most common quantum master equations \cite{Breuer2002}. Luckily, in many cases of practical interest the residual interactions $ \tilde{g}_\mu^{(c)} $ are sufficiently weak so that the dynamics is faithfully captured under this simple assumption. As we show in Sec.~\ref{sec:dynamics} below, this is indeed the case when working in the overdamped limit. Furthermore, given its uniqueness \cite{subasi2012equilibrium}, the non-equilibrium steady state (NESS) of our linear wire is always correctly reproduced by the augmented system, regardless of the initial condition for the RC. 

Before moving on, let us briefly recapitulate: Our original problem consists of two interacting oscillators locally coupled to two heat baths. The coupling to one of them is of the form \eqref{eq:spectral_density_structured}, which complicates the analysis as it is likely to produce non-Markovian dissipation (i.e., with long memory times). Luckily this precise dissipative dynamics can be exactly mimicked by replacing the problematic thermal contact with one auxiliary oscillator undergoing purely Markovian dissipation. In a suitable parameter range, this `augmented' three-oscillator model can thus be tackled via a standard master equation (as we do in Sec.~\ref{sec:GME} below), which would allow us to recover the original dynamics by just tracing out the auxiliary coordinate. The ``twist'' of this paper is that we push such master equation far beyond its range of applicability---namely, we allow for a very strong residual dissipation on the augmented system---and benchmark its prediction for the steady state of the wire against the \textit{exact} stationary solution of the problem. This can always be obtained with the methods outlined in Sec.~\ref{sec:QLE}, since our $ \pmb H $ in Eq.~\eqref{eq:full_Hamiltonian} is fully linear. 

\subsection{Markovian master equation and its stationary solution}\label{sec:GME}

\subsubsection{The (global) GKLS master equation}

We will now outline the derivation of the adjoint quantum master equation for an \textit{arbitrary linear network of $ N $ harmonic nodes, locally coupled to $ M $ baths}. This is a Born--Markov secular master equation \cite{Breuer2002} in the standard Gorini--Kossakowski--Lindblad--Sudarshan (GKLS) form \cite{gorini1976completely,lindblad1976generators}. In the present paper we shall only be interested in applying it to a simple 1D chain of three (and, in Sec.~\ref{sec:dynamics}, also two) harmonic oscillators with heat baths coupled at both ends. Nonetheless, the general equation is of independent interest, as it can be applied to many problems in quantum transport. 

It is important to stress that we treat dissipation \textit{globally}, as opposed to the widespread `local' or `additive' approach \cite{asadian2013heat}. That is, we acknowledge that even if each bath couples \textit{locally} to one node of the network, the ensuing dissipation affects the system \textit{as a whole}, due to the internal interactions. Indeed, the local approach is known to lead to severe physical inconsistencies \cite{joshi2014oscillators,levy2014local,stockburger2016thermodynamic,kolodynski2018adding,maguire2018environmental}. Rigorously, such local equations are only acceptable when understood as either the lowest-order term in a perturbative expansion of a global master equation in the internal coupling strength \cite{PhysRevE.76.031115,trushechkin2016perturbative}, or as a limiting case of a discrete collisional process \cite{barra2015thermodynamic,barra2018smallest,de2018reconciliation}. In any case, addressing dissipation locally is often the only practical way forward in large interacting non-linear open systems---exact diagonalisation of the full many-body Hamiltonian is, otherwise, required. Remarkably, finding, e.g., the NESS, which sets the transport properties of any interacting linear network, with the ``plug-and-play'' stationary solution below [i.e., Eqs.~\eqref{eq:steady_state} and \eqref{eq:steady_heat_GME}] only requires the diagonalisation of the corresponding $ N \times N $ interaction matrix.

The Hamiltonian of a general linear network can be cast as 
\begin{equation}
\pmb H_N = \frac12\,\big(\vec{\pmb X}^\mathsf{T} \mathsf{V}\,\vec{\pmb X} + \vec{\pmb P}^\mathsf{T} \vec{\pmb P}\big),
\label{eq:Hamiltonian_network}
\end{equation}
assuming again that masses are $ \mathsf{M} = \mathbbm{1} $. Here, $ \vec{\pmb X} $ and $ \vec{\pmb P} $ are $ N $-dimensional vectors containing the position and momentum operators of each node, and $ \mathsf{V} $ is real and symmetric. Let $ \mathsf{P} $ be the orthogonal transformation that brings \eqref{eq:Hamiltonian_network} into the diagonal form $ \pmb H_N = \frac12\,\big( \vec{\pmb\eta}^\mathsf{T}\mathsf{\Omega}^2\,\vec{\pmb \eta} + \vec{\pmb \pi}^\mathsf{T}\vec{\pmb \pi} \big) $, where $ \mathsf{\Omega}_{ij} = \Omega_i \delta_{ij} > 0 $ is a diagonal matrix formed of the normal mode frequencies corresponding to the conjugate variables $ \big\lbrace \pmb{\eta}_i, \pmb{\pi}_i \big\rbrace_{i\in\{1,\cdots,N\}} $ (i.e., $ \vec{\pmb\eta} \coloneqq \mathsf{P}^\mathsf{T}\vec{\pmb X} $). 

The standard derivation of a Born--Markov secular master equation \cite{Breuer2002,levy2014local,gonzalez2017testing} now requires to decompose the `system--environment' couplings [in our case, $ \pmb X_i $ for the $ M $ nodes coupled to local baths, as per Eq.~\eqref{eq:hamiltonian_wire-bath}] as eigen-operators of $ \pmb H_N $. That is $ \pmb X_i = \sum_j \pmb L_i(\Omega_j) + \pmb L_i(\Omega_j)^\dagger $ so that $ [\pmb H_N,\pmb L_i(\Omega_j) ] = - \Omega_j \pmb L_i(\Omega_j) $. These non-Hermitian operators, turn out to be simply  
\begin{equation}
\pmb L_i(\Omega_j) = \frac{\mathsf{P}_{ij}}{\sqrt{2\Omega_j}}\,\pmb b_j\,, \qquad \pmb L_i(-\Omega_j) \coloneqq \pmb L_i(\Omega_j)^\dagger\,, 
\label{eq:jump_operators}
\end{equation}
where $ \pmb b_j = \sqrt{\Omega_j/2}\,\big( \pmb\eta_j + i\,\pmb\pi_j/\Omega_j \big) $. With these definitions, the equation of motion for an arbitrary Heisenberg-picture (Hermitian) operator $ \pmb O(t) $ under the Born--Markov and secular approximations reads \cite{Breuer2002}
\begin{widetext}
\begin{multline}
\frac{\text{d}\pmb O(t)}{\text{d}t} = i \left[ \pmb H_N, \pmb O(t) \right] + \sum\nolimits_{i=1}^M\left(\sum\nolimits_{j=1}^N \Gamma_i(\Omega_j)\,\big(\pmb L_i(-\Omega_j)\,\pmb O(t)\,\pmb L_i(\Omega_j) - \frac12\big\lbrace \pmb L_i(-\Omega_j)\,\pmb L_i(\Omega_j),\pmb O(t) \big\rbrace_+ \big) \right.\\
\left. + \sum\nolimits_{j=1}^N \Gamma_i(\Omega_j)\,e^{-\Omega_j/T_j}\,\big( \pmb L_i(\Omega_j)\,\pmb O(t)\,\pmb L_i(-\Omega_j) - \frac12\big\lbrace \pmb L_i(\Omega_j)\,\pmb L_i(-\Omega_j),\pmb O(t) \big\rbrace_+ \big)\right),
\label{eq:GME_general}
\end{multline}
\end{widetext}
with $ \{\cdot,\cdot\}_+ $ denoting anticommutator and decay rates $ \Gamma_i(\Omega_j)\coloneqq 2\,J_i(\Omega_j)\,\big( 1 - e^{-\Omega_j/T_i} \big)^{-1} $, so that $ \Gamma_i(-\Omega_j)/\Gamma_i(\Omega_j) = \exp{(-\Omega_j/T_i)} $, thus reflecting local detailed balance.

The main appeal of Eq.~\eqref{eq:GME_general} is that it is guaranteed to generate a completely positive and trace-preserving dynamics for the system \cite{gorini1976completely,lindblad1976generators}, unlike other frequently used weak-coupling master equations \cite{suarez1992memory,gaspard1999slippage}. Furthermore, under mild ergodicity assumptions, it admits a unique stationary solution \cite{spohn1977algebraic} which, in the case of a single environmental temperature $ T $, is the thermal equilibrium state $ \pmb\varrho_N(t)\propto \exp{(-\pmb H_N/T)} $ \cite{spohn1978entropy}. Importantly, this means that \textit{no renormalisation needs to be done} on the Hamiltonian $ \pmb H_N $ to recover the correct equilibrium state in the high-temperature limit. For that reason, when applying Eq.~\eqref{eq:GME_general} to the three-node augmented system, we take
\begin{equation}
\pmb H_{3} = \pmb H_w + \pmb\delta\pmb H_{w\text{--}c} - \lambda\,\pmb{X}_c\pmb X_\text{RC} + \frac12\left(\omega_0^2\,\pmb X_\text{RC}^2 + \pmb{P}_\text{RC}^2\right)
\label{eq:augmented_Hamiltonian_GME}
\end{equation}
as the system Hamiltonian; i.e, we discard the renormalisation terms $ \pmb\delta\pmb H_{h\text{--}w} $ and $ \pmb\delta\pmb H_\text{RC--res} $ in Eq.~\eqref{eq:full_Hamiltonian}, corresponding to the thermal contact with the hot and the residual environment, respectively. 

However, the term $ \pmb\delta\pmb H_{w\text{--}c} $ is---by construction---part of the augmented system after the reaction-coordinate mapping \cite{iles2014environmental,strasberg2016nonequilibrium}. As we shall see in Sec.~\ref{sec:dynamics} below, disregarding this latter term in the augmented-system Hamiltonian, e.g., on the basis of $ \delta_c $ being small, can yield the wrong dynamics for the wire at intermediate times, even if the short-time evolution and the steady state are reproduced accurately.

\subsubsection{Equations of motion for the covariances}

Applying Eq.~\eqref{eq:GME_general} to the symmetrised covariances $ \big\langle\frac12 \{ \pmb r_j(t),\pmb r_k(t) \}_+\big\rangle \coloneqq [\mathsf{c}_\text{me}^{(N)}]_{jk}(t) $, where $ \vec{\pmb r} = (\pmb\eta_1,\pmb\pi_1,\cdots,\pmb\eta_N,\pmb\pi_N)^\mathsf{T} $, yields a closed algebra for the `covariance matrix' of the network $ \mathsf{c}_\text{me}^{(N)}(t) $, where the sub-index `me' stands for `master equation' and allows to differentiate it from the `ex' (for `exact') covariance matrix, that we will compute in Sec.~\ref{sec:QLE} below. Specifically, we have
\begin{widetext}
\begin{subequations}
\begin{align}
\frac{\text{d}}{\text{d}t}\langle\pmb\eta_j^2\rangle &= \sum\nolimits_{i=1}^M \frac{\mathsf{P}_{ij}^2}{2\Omega_j}\,\Delta_i(\Omega_j)\,\langle\pmb\eta_j^2\rangle + \langle\{\pmb\eta_j,\pmb\pi_j \}_+\rangle + \sum\nolimits_{i=1}^M \frac{\mathsf{P}_{ij}^2}{4\Omega_j^2}\,\Sigma_i(\Omega_j) \\
\frac{\text{d}}{\text{d}t} \langle \{ \pmb\eta_j,\pmb\pi_j \}_+ \rangle &= - 2\Omega_j^2\langle\pmb\eta_j^2\rangle + 2\langle\pmb\pi_j^2\rangle + \sum\nolimits_{i=1}^M \frac{\mathsf{P}_{ij}^2}{2\Omega_j}\,\Delta_i(\Omega_j)\,\langle \{ \pmb\eta_j,\pmb\pi_j \}_+\rangle \\
\frac{\text{d}}{\text{d}t}\langle\pmb\pi_j^2\rangle &= \sum\nolimits_{i=1}^M\frac{\mathsf{P}_{ij}^2}{2\Omega_j}\Delta_i(\Omega_j)\,\langle \pmb\pi_j^2\rangle - \Omega_j^2 \langle\{\pmb\eta_j,\pmb\pi_j \}_+\rangle + \sum\nolimits_{i=1}^M \frac{\mathsf{P}_{ij}^2}{4}\,\Sigma_i(\Omega_j),
\end{align} 
\label{eq:GME_main_covariances}
\end{subequations}
together with the asymptotically vanishing covariances (for $ j \neq k $)
\begin{subequations}
\begin{align}
\frac{\text{d}}{\text{d}t}\langle\pmb\eta_j\,\pmb\eta_k\rangle &= \langle \pmb\eta_j\,\pmb\pi_k \rangle + \langle \pmb\eta_k\,\pmb\pi_j \rangle + \sum\nolimits_{i=1}^M \left(\frac{\mathsf{P}_{ij}^2}{4\Omega_j}\Delta_i(\Omega_j) + \frac{\mathsf{P}_{ik}^2}{4\Omega_k}\Delta_i(\Omega_k)\right)\langle\pmb\eta_j\,\pmb\eta_k\rangle \\
\frac{\text{d}}{\text{d}t}\langle\pmb\eta_j\,\pmb\pi_k\rangle &= \langle\pmb\pi_j\,\pmb\pi_k\rangle - \Omega_k^2\langle\pmb\eta_j\,\pmb\eta_k\rangle + \sum\nolimits_{i=1}^M\left( \frac{\mathsf{P}_{ij}^2}{4\Omega_j}\Delta_i(\Omega_j) + \frac{\mathsf{P}_{ik}^2}{4\Omega_k}\Delta_i(\Omega_k) \right)\langle\pmb\eta_j\,\pmb\pi_k\rangle \\
\frac{\text{d}}{\text{d}t}\langle\pmb\pi_j\,\pmb\pi_k\rangle &= -\Omega_j^2\langle\pmb\eta_j\,\pmb\pi_k\rangle-\Omega_k^2\langle\pmb\eta_k\,\pmb\pi_j\rangle + \sum\nolimits_{i=1}^M\left( \frac{\mathsf{P}^2_{ij}}{4\Omega_j}\Delta_i(\Omega_j) + \frac{\mathsf{P}_{ik}^2}{4\Omega_k}\Delta_i(\Omega_k) \right) \langle\pmb\pi_j\,\pmb\pi_k\rangle, 
\end{align}
\label{eq:GME_other_covariances}
\end{subequations}
\end{widetext}
where $ \Sigma_i(\Omega_j) \coloneqq \Gamma_i(-\Omega_j) + \Gamma_i(\Omega_j) $ and $ \Delta_i(\Omega_j)\coloneqq \Gamma_i(-\Omega_j)-\Gamma_i(\Omega_j) $. For completeness, the equations of motion for the first-order moments $ \langle\pmb\eta_j\rangle $ and $ \langle\pmb\pi_j\rangle $ are given by
\begin{subequations}
\begin{align}
\frac{\text{d}}{\text{d}t} \langle\pmb\eta_j\rangle &= \langle\pmb\pi_j\rangle + \sum\nolimits_{i=1}^M\frac{\mathsf{P}_{ij}^2}{4\Omega_j} \Delta_i(\Omega_j)\,\langle\pmb\eta_j\rangle \\
\frac{\text{d}}{\text{d}t}\langle\pmb\pi_j\rangle &= -\Omega_j^2\langle\pmb\eta_j\rangle + \sum\nolimits_{i=1}^M\frac{\mathsf{P}_{ij}^2}{4\Omega_j}\Delta_i(\Omega_j)\,\langle\pmb\pi_j\rangle\,.
\end{align}
\label{eq:GME_first-order}
\end{subequations}

Since our Hamiltonian \eqref{eq:Hamiltonian_network} is quadratic in position and momenta, any Gaussian initial state of the network will remain Gaussian at all times. In turn, given that Gaussian states are fully characterised by their first- and second-order moments \cite{0503237v1} (that is, $ \langle\pmb r_j(t)\rangle $ and $ \big\langle\frac12\{ \pmb r_j(t),\pmb r_k(t) \}_+ \big\rangle $), Eqs.~\eqref{eq:GME_main_covariances}--\eqref{eq:GME_first-order} thus provide a full dynamical description of the problem. Furthermore, since $ \langle \pmb r_j(\infty) \rangle = \langle \pmb r_j(\infty)\,\pmb r_k(\infty) \rangle = 0 $ for $ j\neq k $, we can concentrate only in Eqs.~\eqref{eq:GME_main_covariances} as far as the NESS is concerned. Explicitly, this is given by
\begin{subequations}
\begin{align}
&\langle \pmb\eta^2_j(\infty) \rangle = -\frac{\tilde{\Sigma}(\Omega_j)}{2 \tilde{\Delta}(\Omega_j)\,\Omega_j},\\
&\big\langle \frac12 \{ \pmb\eta_j(\infty),\pmb\pi_j(\infty) \}_+\big\rangle = 0,\\ 
&\langle\pmb\pi_j^2(\infty)\rangle = - \frac{\tilde{\Sigma}(\Omega_j)\,\Omega_j}{2\tilde{\Delta}(\Omega_j)},
\end{align}
\label{eq:steady_state}
\end{subequations}
where $ \tilde{\Sigma}(\Omega_j) \coloneqq \sum\nolimits_{i=1}^M \mathsf{P}_{ij}^2/(2\Omega_j)\,\Sigma_i(\Omega_j) $ and $ \tilde{\Delta}(\Omega_j) \coloneqq \sum\nolimits_{i=1}^M \mathsf{P}_{ij}^2/(2\Omega_j)\,\Delta_i(\Omega_j) $. One can then transform $ \mathsf{c}^{(N)}(t) $ into the covariance matrix $ \mathsf{C}^{(N)}_\text{me}(t) $, defined in terms of the original variables $ \vec{\pmb R} = (\pmb X_1, \pmb P_1, \cdots, \pmb X_N, \pmb P_N )^\mathsf{T} $ by means of $ \mathsf{C}_\text{me}^{(N)}(t) = \mathsf{Q}\,\mathsf{c}_\text{me}^{(N)}(t)\,\mathsf{Q}^\mathsf{T} $, where
\begin{equation}
\mathsf{Q} = \left(
\begin{array}{ccccc}
\mathsf{P}_{11} & 0 & \mathsf{P}_{12} & 0 & \cdots \\
0 & \mathsf{P}_{11} & 0 & \mathsf{P}_{12} & \cdots \\
\mathsf{P}_{21} & 0 & \mathsf{P}_{22} & 0 & \cdots \\
0 & \mathsf{P}_{21} & 0 & \mathsf{P}_{22} & \cdots \\
\vdots & \vdots & \vdots & \vdots & \ddots
\end{array}
\right)\,.
\end{equation} 

{\color{markedgreen}Importantly, under the secular approximation underpinning the GKLS equation, all the position--momentum covariances $ \langle\pmb{X}_i(\infty)\,\pmb{P}_j(\infty)\rangle $ vanish in steady state. As a result, local current operators defined within the harmonic network would invariably average to zero \cite{PhysRevE.76.031115}. We shall elaborate more on this in Sec.~\ref{sec:steady_state_heat} below. In order to compute the correct stationary heat currents, one can alternatively} define the adjoint dissipation super-operators $ \mathcal{L}_i^\dagger $ for each heat bath by rewriting Eq.~\eqref{eq:GME_general} as $ \text{d} \pmb O(t) / \text{d}t \coloneqq i[\pmb H_N,\pmb O] + \sum_{i=1}^M \mathcal{L}_i^\dagger\,\pmb O $. That way, we can cast the steady-state heat current flowing from the $i$th bath into the network as $ \dot{\mathcal{Q}}_{i,\text{me}}^{(N)} \coloneqq \langle \mathcal{L}_i^\dagger\,\pmb H_N(\infty) \rangle $ \cite{alicki1979engine,Kosloff2014}. In our case, this evaluates to 
\begin{multline}
\dot{\mathcal{Q}}_{i,\text{me}}^{(N)} = \sum\nolimits_{j=1}^N \frac{\mathsf{P}_{ij}^2}{2\Omega_j}\,\Delta_i(\Omega_j) \left( \frac12\Omega_j^2\langle\pmb\eta_j^2(\infty)\rangle + \frac12\langle\pmb\pi_j^2(\infty)\rangle \right) \\ + \frac{\mathsf{P}_{ij}^2}{4}\,\Sigma_i(\Omega_j)\,.
\label{eq:steady_heat_GME}
\end{multline}

In Sec.~\ref{sec:steady-state-and-heat-currents} below, we shall apply the general equations \eqref{eq:steady_state} and \eqref{eq:steady_heat_GME} to the simple three-oscillator chain making up the augmented system for our quantum wire (cf. Fig.~\ref{fig:0}), and compare them with the \textit{exact} stationary state and heat currents (see Sec.~\ref{sec:QLE}). In turn, in Sec.~\ref{sec:dynamics}, we compare the reduced \textit{dynamics} of the augmented system with the time-evolution of the two-node wire in a parameter regime where Eqs.~\eqref{eq:GME_main_covariances}--\eqref{eq:GME_first-order} are also directly applicable to the original problem.

\subsubsection{A note on the underlying approximations} \label{sec:GKLS_approximations}

To conclude this section, let us briefly comment on the approximations underlying the microscopic derivation of Eq.~\eqref{eq:GME_general} \cite{Breuer2002}. First and foremost, it is a second-order perturbative expansion of the exact master equation in the system--environment(s) coupling \cite{gaspard1999slippage}. Therefore, \textit{it is only meaningful under the assumption of weak dissipation}. In addition, the Markov approximation has been performed by neglecting any memory effects in the dissipative process, since environmental correlations are assumed to be very short-lived. Note that it may well be the case that environmental correlations are indeed short while the dissipation is strong; recall that the bath memory time is essentially determined by the ``shape'' of the spectral density [cf. Eq.~\eqref{eq:environmental-correlation}]. In such situation, the Markov approximation would be valid, but the weak coupling assumption would be violated.

The completely positive GKLS form \eqref{eq:GME_general} is attained after performing the secular approximation which, in our case requires that all normal-mode frequencies $ \Omega_j $ be well separated as compared to the dissipation rates (i.e., $ \min_{j\neq k}\{ \vert \Omega_j-\Omega_k \vert, 2 \Omega_j \} \gg \max_{i\in\{1,\cdots,M\}}\,\{\gamma_i\} $). Once again, this approximation is incompatible with arbitrarily large dissipation rates $ \gamma_i $, but may also be easily violated under weak dissipation \cite{PhysRevE.76.031115,gonzalez2017testing}. For that reason, the full Redfield equation \cite{suarez1992memory,gaspard1999slippage}---containing all non-secular terms---is often used instead when performing the RC mapping \cite{iles2014environmental,iles2016energy,strasberg2016nonequilibrium}. As we will see in Sec.~\ref{sec:discussion} below, even if \textit{both} the weak coupling and the secular approximation are violated on the augmented system, the two-node reduction of the resulting state may still provide an excellent approximation to the exact steady state of the wire. 

As a final remark, notice that Eq.~\eqref{eq:GME_general} does not include the so-called Lamb shift term \cite{Breuer2002}. This is a Hamiltonian-like contribution to the master equation, dissipative in origin. The Lamb shift is often neglected for being a `small' contribution when compared with the bare Hamiltonian $ \pmb H_N $ \cite{strasberg2016nonequilibrium}. It is safe to say that, when working with a GKLS quantum master equation, the Lamb shift is entirely irrelevant for the thermodynamics of \textit{steady-state} energy-conversion processes \footnote{Indeed, all the standard quantum-thermodynamic arguments based on the contractivity of the dissipative dynamics and uniqueness of the (thermal) fixed point of any of the dissipators of a GKLS equation \cite{spohn1978entropy,alicki1979engine} hold regardless of whether or not the Lamb shift is included in $ \mathcal{L}_i $. In particular, one \textit{always} finds $ \mathcal{L}_i\,e^{-\pmb H_N/T_i} = 0 $, i.e., the fixed point of the dynamics is a thermal state with respect to the \textit{unshifted} Hamiltonian.}. Interestingly, however, when the Redfield equation is used instead (e.g., due to the inadequacy of the secular approximation), the Lamb shift can have noticeable effects \cite{strasberg2018fermionic}. Note that this term is not related to the frequency renormalisation discussed in Sec.~\ref{sec:full_hamiltonian} above. 

\subsection{Exact stationary solution}\label{sec:QLE}

The stationary state of our two-node wire can be obtained \textit{exactly}, with no other assumptions than a factorised initial state of the form $ \pmb\rho(0) = \pmb\varrho_{T_h}\otimes\pmb\varrho_w(0)\otimes\pmb\varrho_{T_c} $, and no restrictions on $ \pmb\varrho_w(0) $. Importantly, the problem can be solved analytically \textit{regardless of the spectral densities} at the boundaries. These linear open systems have been extensively studied in the literature \cite{riseborough1985exact,ludwig2010entanglement,fleming2011exact,correa2012asymptotic,martinez2013dynamics,valido2015currents,freitas2014analytic}, as they are among the few which admit an exact solution under strong dissipation. Full details about the calculation of the steady state and stationary heat currents for the Hamiltonian in Eq.~\eqref{eq:original_hamiltonian} were given by Gonz\'alez \textit{et al.} \cite{gonzalez2017testing}, and here we limit ourselves to outline the key steps.

The exact dynamics of the wire obeys the following quantum Langevin equations \cite{ford1965statistical,ford1988qle}
\begin{multline}
\frac{\text{d}^2}{\text{d}t^2}\pmb X_\alpha + (\omega_\alpha^2 + \delta_\alpha)\,\pmb X_\alpha + k\,(\pmb X_\alpha-\pmb X_{\bar{\alpha}})  \\= \pmb F_\alpha(t) + \int_{t_0}^{t}\text{d}s\,\chi_\alpha(t-s)\,\pmb X_\alpha(s), 
\label{eq:QLE}
\end{multline}
where $ \alpha\in\{h,c\} $ and $ \bar{c} \coloneqq h $ and $ \bar{h} \coloneqq c $. As we can see, the coherent evolution of the two coupled (and renormalised) oscillators is affected by environmental driving and dissipation (terms on the right-hand side). Importantly, the upper limit of the integral can be extended to infinity by supplementing the dissipation kernel $ \chi_\alpha(t) $ with a Heavisde step function $ \Theta(t) $ [i.e., $ \chi_\alpha(t)\mapsto\chi_\alpha(t)\,\Theta(t) $] \cite{correa2017strong}. Since we are interested in the steady state of the wire, our aim will be to compute the covariance matrix $ \mathsf{C}^{(2)}_\text{ex} $ at any finite time $ t $ while setting $ t_0\rightarrow-\infty $. 

With this in mind, we can now Fourier-transform Eqs.~\eqref{eq:QLE}, which yields
\begin{widetext}
\begin{equation}
\left(
\begin{array}{cc}
-\omega^2 + \omega_h^2 + \delta_h + k - \hat{\chi}_h(\omega) & -k \\
-k & -\omega^2 + \omega_c^2 + \delta_c + k - \hat{\chi}_c(\omega)
\end{array}
\right) \left(\begin{array}{c}
\hat{\pmb X}_h \\
\hat{\pmb X}_c
\end{array}\right) 
\coloneqq \mathsf{A}(\omega)\left(\begin{array}{c}
\hat{\pmb X}_h \\
\hat{\pmb X}_c
\end{array}\right)
= 
\left(
\begin{array}{c}
\hat{\pmb F}_h \\
\hat{\pmb F}_c
\end{array}
\right)\,.
\label{eq:QLE_fourier}
\end{equation}
Here, the ``hatted'' symbols are in the frequency domain, i.e., $ \hat{f}(\omega) \coloneqq \int_{-\infty}^{\infty}\text{d}t\, e^{i\omega t}\,f(t) $. Therefore, $ (\hat{\pmb X}_h,\hat{\pmb X}_c)^\mathsf{T} = \mathsf{A}^{-1}(\omega)\,(\hat{\pmb F}_h, \hat{\pmb F}_c)^\mathsf{T} $, so that the objects we wish to compute are
\begin{equation}
\big\langle\frac12\{\pmb X_\alpha(t'),\pmb X_\beta(t'')\} \big\rangle = \int_{-\infty}^\infty\frac{\text{d}\omega'}{2\pi}\int_{-\infty}^\infty\frac{\text{d}\omega''}{2\pi}\,\sum\nolimits_{\delta\gamma}[\mathsf{A}^{-1}]_{\alpha\gamma}(\omega')\,[\mathsf{A}^{-1}]_{\,\beta\delta}(\omega'')
\,\big\langle \frac12\{\hat{\pmb F}_\gamma(\omega'),\hat{\pmb F}_\delta(\omega'')\} \big\rangle\,e^{-i\omega' t'} e^{-i\omega''t''}
\label{eq:inverse_Fourier_covariances}
\end{equation} 
\end{widetext}
for $ t' = t'' = t $. The position--momentum and momentum--momentum covariances can be obtained by differentiating Eq.~\eqref{eq:inverse_Fourier_covariances}, which is equivalent to multiplying the integrand by $ (-i\omega') $ and $ (-\omega'\omega'') $, respectively. To carry out the integration in \eqref{eq:inverse_Fourier_covariances} explicitly, we only need the Fourier transform of the dissipation kernels $ \hat{\chi}_\alpha(\omega) $ and the power spectrum of the environmental forces $ \langle \frac12\{\hat{\pmb F}_\alpha(\omega'),\hat{\pmb F_\beta}(\omega'') \}\rangle $. These are given by \cite{correa2017strong}
\begin{subequations}
\begin{align}
&\text{Im}\,\hat{\chi}_\alpha = J_\alpha(\omega)\,\Theta(\omega) - J_\alpha(-\omega)\,\Theta(-\omega) \label{eq:Im_chi}\\
&\text{Re}\,\hat{\chi}_\alpha = \frac{1}{\pi}\,\text{P}\int_{-\infty}^\infty\text{d}\omega'\,\frac{\text{Im}\,\hat{\chi}_\alpha(\omega')}{\omega'-\omega} \label{eq:Re_chi}\\
&\big\langle \frac12\{ \hat{\pmb F}_\alpha(\omega'),\hat{\pmb F}_\beta(\omega'') \} \big\rangle = \frac{1}{2\pi}\,\coth{\Big( \frac{\omega'}{2T_\alpha} \Big)}\,\text{Im}\,\hat{\chi}_\alpha\nonumber\\
& \qquad\qquad\qquad\qquad\qquad\qquad\times\delta_{\alpha\beta}\,\delta(\omega'+\omega'') \label{eq:noise_power_spectrum}\,,
\end{align}
\label{eq:missing_pieces_QLE}
\end{subequations}
where $ \text{P} $ denotes `principal value', $ \delta_{\alpha\beta} $ is a Kronecker delta and $ \delta(x) $ is a Dirac delta. The integration in Eq.~\eqref{eq:Re_chi} can be readily performed for the overdamped and underdamped spectral densities of interest, i.e., $ J_h(\omega) =  \gamma_h\Lambda_h^2\omega/(\omega^2 + \Lambda_h^2)$ and $ J_c(\omega) = \gamma\lambda^2\omega/[\gamma^2\omega^2 + (\omega^2-\omega_0^2)^2] $, which yields
\begin{subequations}
\begin{align}
\hat{\chi}_h(\omega) &= \frac{\Lambda_h^2\,\gamma_h}{\Lambda_h-i\omega} \label{eq:chi_h}\\
\hat{\chi}_c(\omega) &= \frac{\lambda^2}{\omega_0^2-i\gamma\omega-\omega^2}\label{eq:chi_c}\,.
\end{align}
\label{eq:Fourier_dissipation_kernels}
\end{subequations}
Note that Eq.~\eqref{eq:chi_h} may also be used for the dissipation kernel of the residual bath acting on the augmented system, by merely replacing $ \gamma_h $ with $ \gamma $ and taking a large cutoff.

Summing up, Eqs.~\eqref{eq:inverse_Fourier_covariances}--\eqref{eq:Fourier_dissipation_kernels} are all we need to fill in the full stationary $ 4\times 4 $ covariance matrix $ \mathsf{C}^{(2)}_\text{ex}(\infty) $. Note that it is indeed possible to solve the problem not only exactly, but also analytically \cite{correa2012asymptotic,PhysRevA.88.012309}. In turn, the $ 6\times 6 $ NESS $ \mathsf{C}^{(3)}_\text{ex}(\infty) $ of the augmented system can be found in a completely analogous way \cite{PhysRevA.88.012309}, by just replacing the `vector of forces' $ (\hat{\pmb F}_h,\hat{\pmb F}_c)^\mathsf{T} $ by $ (\hat{\pmb F}_h,0,\hat{\pmb F}_\text{res})^\mathsf{T} $ and $ \mathsf{A}(\omega) $, with  
\begin{widetext}
\begin{equation}
\mathsf{B}(\omega) = \left(
\begin{array}{ccc}
-\omega^2 + \omega_h^2 + \delta_h + k - \hat{\chi}_h(\omega) & -k & 0\\
-k & -\omega^2 + \omega_c^2 + \delta_c + k & -\lambda \\
0 & -\lambda & -\omega^2 + \omega_0^2 + \delta_\text{res} - \hat{\chi}_\text{res}(\omega)
\end{array}
\right)\,.
\label{eq:susceptibility_matrix_augmented}
\end{equation}
\end{widetext}

To conclude this section, let us introduce the exact stationary heat currents, for comparison with Eq.~\eqref{eq:steady_heat_GME}. A direct calculation shows that the change in the energy of our wire (or the augmented system) due to dissipative interactions with bath $ \alpha $---i.e, $ \dot{\mathcal{Q}}^{(N)}_{\alpha,\text{ex}} = i \langle [\pmb{H}_{w\text{--}\alpha},\pmb{H}_w] \rangle $---can be cast as \cite{martinez2013dynamics,freitas2017fundamental}
\begin{subequations}
\begin{align}
\dot{\mathcal{Q}}_{h,\text{ex}}^{(2)} &= -k \langle\pmb X_c\pmb P_h\rangle_{N=2} = k \langle \pmb X_h \pmb P_c \rangle_{N=2} = -\dot{\mathcal{Q}}_{c,\text{ex}}^{(2)} \label{eq:exact_stationary_heat_2} \\
\dot{\mathcal{Q}}_{h,\text{ex}}^{(3)} &= -k \langle\pmb X_c\pmb P_h\rangle_{N=3} = \lambda\langle\pmb X_c \pmb P_\text{RC} \rangle_{N=3} = -\dot{\mathcal{Q}}_\text{res,ex}^{(3)}\,.
\end{align}
\label{eq:exact_stationary_heat}
\end{subequations}

\section{Discussion}\label{sec:discussion}

\subsection{Steady state and stationary heat currents}\label{sec:steady-state-and-heat-currents}

\begin{figure}[t!]
	\centering
	\includegraphics[width=0.83\linewidth]{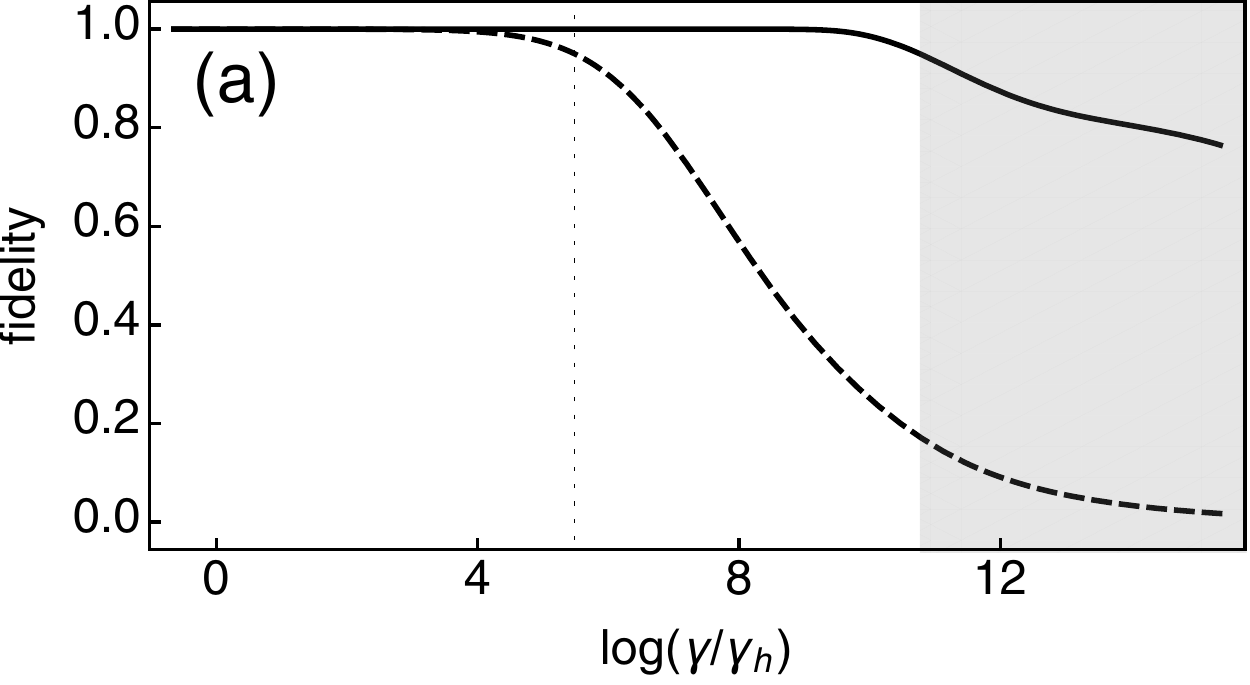}
	\includegraphics[width=0.83\linewidth]{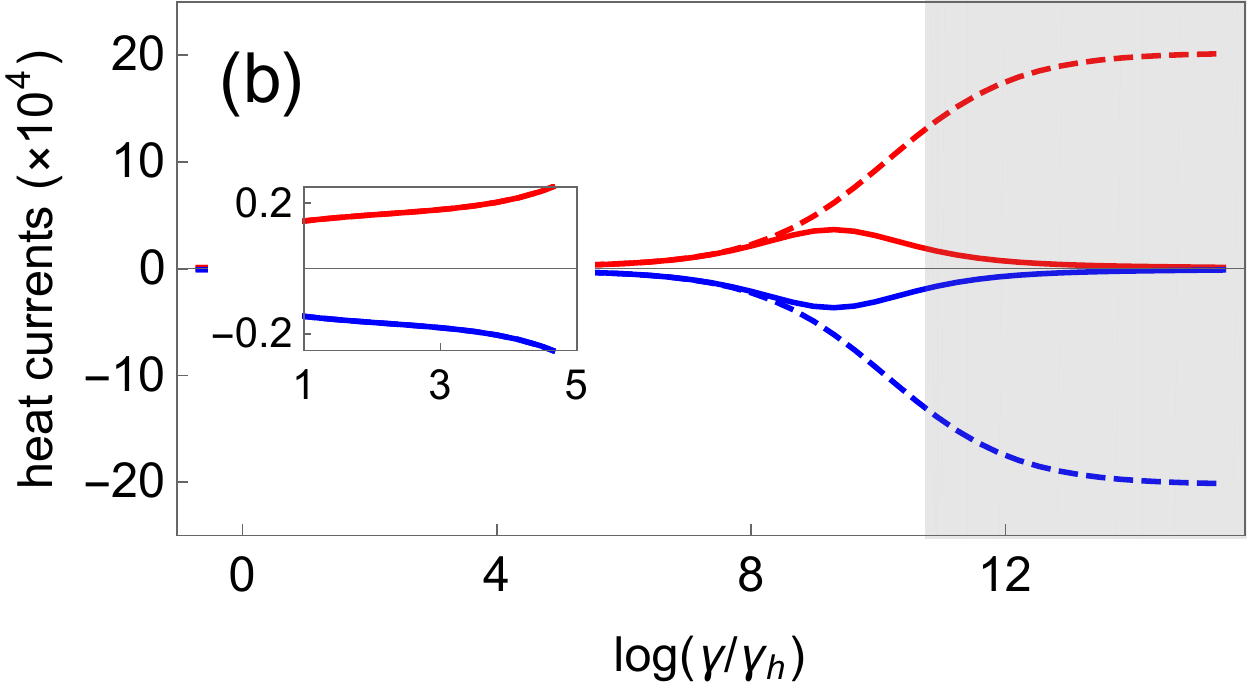}
	\caption{\textbf{(a)} (solid) Uhlmann fidelity between $ \mathsf{C}_\text{ex}^{(2)}(\infty) $ and the relevant two-node reduction of $ \mathsf{C}_\text{me}^{(3)}(\infty) $. This is achieved simply by eliminating rows and columns related to the reaction-coordinate variables $ \pmb X_\text{RC} $ and $ \pmb P_\text{RC} $ from the $ 6\times 6 $ matrix $ \mathsf{C}_\text{me}^{(3)}(\infty) $. The abscissa corresponds to the {\color{markedgreen}natural} $ \log $ of the friction coefficient $ \gamma $ of the underdamped spectral density in Eq.~\eqref{eq:spectral_density_structured}, at the interface between the wire and the cold bath (normalised by the dissipation strength into the hot bath $ \gamma_h $). Only in the shaded-grey area the fidelity drops below $ 95\% $. (dashed) Fidelity between $ \mathsf{C}_\text{ex}^{(3)}(\infty) $ and $ \mathsf{C}_\text{me}^{(3)}(\infty) $ as a function of the normalised friction $ \gamma/\gamma_h $. {\color{markedgreen}This falls below 95\% to the right of the dotted line}. \textbf{(b)} Steady-state heat currents coming from the hot (red) and cold (blue) baths \textit{into} the wire as a function of $ \gamma/\gamma_h $ for the same parameters as in panel (a). The solid lines correspond to the exact calculation from Eq.~\eqref{eq:exact_stationary_heat_2}, while the dashed ones result from applying the GKLS master equation to the augmented system, as per Eq.~\eqref{eq:steady_heat_GME}. The inset is a zoom into the low friction limit and the shaded-grey area is the same as in (a). In both panels, the wire is characterised by $ \omega_h = 1 $, $ \omega_c = 3 $, and $ k = 0.8 $; the overdamped Ohmic spectral density at the hot interface, by $ \gamma_h = 10^{-3} $ and $ \Lambda_h = 10^3 $; the underdamped spectral density at the cold interface, by $ \lambda = 0.9 $ and $ \omega_0 = 4 $; and the baths, by temperatures $ T_h = 3.3 $ and $ T_c = 1.2 $.}
\label{fig:1}
\end{figure}

We are now ready to put the reaction-coordinate mapping to the test. Using Eqs.~\eqref{eq:inverse_Fourier_covariances}--\eqref{eq:Fourier_dissipation_kernels}, we can compute the exact stationary covariance matrix of the original (two-node wire) problem $ \mathsf{C}^{(2)}_\text{ex}(\infty) $, as well as that of the augmented (three-node) system, $ \mathsf{C}^{(3)}_\text{ex}(\infty) $. Alternatively, we can look for the steady state of the augmented system according to the GKLS master equation, i.e., $ \mathsf{C}_\text{me}^{(3)} $. Benchmarking the RC mapping thus amounts to assessing how ``close'' is the relevant $ 4 \times 4 $ submatrix of $ \mathsf{C}_\text{me}^{(3)}(\infty) $ to the exact stationary state $ \mathsf{C}_\text{ex}^{(2)}(\infty) $. {\color{markedgreen} We conclude by noting that the covariance dynamics can also be obtained non-perturbatively in the system--baths couplings by means of stochastic propagation and averaging, in linear and weakly non-linear continuous-variable systems \cite{motz2017currents,motz2018rectification}.}

We thus need to be able to \textit{quantify} the distance between two covariance matrices $ \mathsf{C}_1 $ and $ \mathsf{C}_2 $. To that end, we resort to the Uhlmann fidelity \cite{uhlmann1976transition,banchi2005fidelity} $ \mathcal{F}(\mathsf{C}_1,\mathsf{C}_2) $ which, for arbitrary $N$-mode Gaussian states with vanishing first-order moments, is
\begin{subequations}
\begin{align}
&\mathcal{F}(\mathsf{C}_1,\mathsf{C}_2) = \left(\frac{F}{\sqrt[4]{\text{det}\,\left(\mathsf{C}_1 + \mathsf{C}_2\right)}}\right)^2 \\
&F \coloneqq \left(\text{det}\,\left( 2 \left(\sqrt{\mathbbm{1} + \frac{\big( \mathsf{C}_\text{aux}\,\mathsf{\Theta} \big)^{-2}}{4}} + \mathbbm{1}\right)\,\mathsf{V}_\text{aux} \right)\right)^{1/4} \\
&\mathsf{C}_\text{aux} \coloneqq \mathsf{\Theta}^\mathsf{T}\left( \mathsf{C}_1 + \mathsf{C}_2 \right)^{-1}\left( \frac{\mathsf{\Theta}}{4} + \mathsf{C}_2\,\mathsf{\Theta}\,\mathsf{C}_1 \right) \\
&\mathsf{\Theta} \coloneqq \bigoplus\nolimits_{i=1}^N \left(\begin{array}{cc}
0 & 1 \\
-1 & 0
\end{array}\right).
\end{align}
\label{eq:uhlmann}
\end{subequations}
This is a meaningful distance measure, since $ \mathcal{F}(\mathsf{C}_1,\mathsf{C}_2) = 1 $ only holds if the states are identical, and $ 0 < \mathcal{F}(\mathsf{C}_1,\mathsf{C}_2) \leq 1 $.

\subsubsection{Steady states}\label{sec:steady_states}

In Fig.~\hyperref[fig:1]{2(a)} we illustrate our steady-state benchmark for the RC mapping (solid line). Strikingly, we find that the reduction of $ \mathsf{C}_\text{me}^{(3)}(\infty) $ onto the wire degrees of freedom remains nearly identical to the exact stationary state $ \mathsf{C}_\text{ex}^{(2)}(\infty) $, even at extremely large residual dissipation strengths $ \gamma $. In the figure, for instance, the fidelity between the two states falls below $ 95 \% $ only at $ \gamma \gtrsim 60 $. When it comes to the approximations that justify the GKLS master equation \eqref{eq:GME_general} (cf. Sec.~\ref{sec:GKLS_approximations}), this is \textit{completely off-limits}. Indeed, note that the normal-mode frequencies of the augmented system are, in this example, $ \Omega_1 \simeq 1.31 $, $ \Omega_2 \simeq 3.13 $, and $ \Omega_3 \simeq 4.02 $, which renders the secular approximation problematic already at residual dissipations as small as $ \gamma \sim 0.1 $. More importantly, $ \gamma \simeq 60 $ can by no means be considered \textit{small} and hence, a perturbative expansion of the generator of the dissipative dynamics is out of the question. Our extensive numerics show that this surprising observation is not due to a lucky parameter choice but rather, a \textit{generic feature}. It is also consistent with the excellent agreement previously reported in other (non-linear) models \cite{iles2014environmental,iles2016energy,lambert2019virtual}, between the reduction of the master-equation-propagated augmented system and the numerical solution to the original problem. 

As surprising as this observation may seem, there is nothing contradictory in it---indeed, the GKLS master equation does break down for $ \gamma \gtrsim 0.1 $, which corresponds to $ \log{(\gamma/\gamma_h)} \gtrsim 5 $ {\color{markedgreen}[area to the right of the dotted line in Fig.~\ref{fig:1}(a)]}. We can see this in Fig.~\hyperref[fig:1]{2(a)}, when instead of looking at the \textit{reduction} of $ \mathsf{C}_\text{me}^{(3)}(\infty) $ onto the wire, we consider the \textit{full} augmented system and compare it with the exact three-node solution $ \mathsf{C}_\text{ex}^{(3)}(\infty) $ (dashed line). Specifically, $ \mathcal{F}(\mathsf{C}_\text{ex}^{(3)}(\infty),\mathsf{C}_\text{me}^{(3)}(\infty)) < 0.95 $ for $ \gamma > 0.15 $, as expected. We are thus not claiming that Markovian master equations in Lindblad form are generally valid for strong coupling situations. What we find is that non-equilibrium energy transfer processes through open quantum systems in complex environments can be captured faithfully over a much wider parameter range than previously thought, by combining the RC mapping with a GKLS master equation (RC--GKLS mapping). 

{\color{markedgreen} We still need, however, to provide some physical intuition backing this observation. To that end, let us take a \textit{detour} to comment on recent literature on \textit{locality of temperature} in quantum many-body lattice systems \cite{ferraro2012intensive,PhysRevA.79.052340,PhysRevX.4.031019,hernandez2015locality}. It is clear that the reduction of the global thermal state of a large lattice onto a small local subspace can deviate substantially from a local thermal state---this is due to the non-vanishing interactions between the subsystem in question and the rest of the lattice. However, the (non-thermal) state of such sub-lattice may be approximated arbitrarily well as follows: One first envelopes it with a `boundary' or `buffer' region, taken from the surrounding lattice; such augmented system is then set to a thermal state at the global temperature of the full system and then, the auxiliary buffer is traced out \cite{ferraro2012intensive}. The result is in good agreement with the local state of interest so long as the boundary is thick enough, relative to some relevant correlation length scale \cite{PhysRevX.4.031019,hernandez2015locality}. Something similar happens in our example: imposing incorrect (thermal) boundary conditions on an augmented system, we can reproduce the state of the wire faithfully; the techniques only breaks down when the boundary--environment interactions become sufficiently large, so that correlations start to appear between the wire and the residual environment. Making this intuition more precise by studying the correlation sharing structure between wire, RC, and residual bath, goes, however, beyond the scope of the present paper.}

\subsubsection{Steady-state heat currents}\label{sec:steady_state_heat}

Besides faithfully reproducing the NESS of an open quantum system, one would also like to learn about the stationary heat currents that it supports, especially when viewing it as a `continuous thermal device' for quantum thermodynamics \cite{Kosloff2014}. To do so from the RC-mapped picture, we need to gauge the energy per unit time crossing the boundary between either bath and the augmented system; this can only be achieved by using to the corresponding GKLS dissipators $ \mathcal{L}_i $ [cf. Eq.~\eqref{eq:steady_heat_GME}]. Under strong coupling, however, these are certainly \textit{not} valid generators of the dissipative dynamics. \textit{A priori}, one should thus expect a substantial mismatch between the GKLS stationary heat currents and their exact values in this regime. In Fig.~\hyperref[fig:1]{2(b)} we can indeed see that for $ \gamma \sim 60 $---where $ \mathsf{C}_\text{ex}^{(2)}(\infty) $ and the reduction of $ \mathsf{C}_\text{me}^{(3)}(\infty) $ differ only by $ 5\% $---the master equation overestimates the heat currents by an order of magnitude, and fails to capture, even qualitatively, their behaviour for larger friction $ \gamma $. 

Note that, for us, resorting to the dissipators is indeed the only feasible way to estimate heat currents; $ \mathsf{C}_\text{me}^{(3)}(\infty) $ is lacking the key covariances $ \langle \pmb{X}_c\,\pmb{P}_h \rangle $ and $ \langle \pmb{X}_c\,\pmb{P}_\text{RC} \rangle $ needed to evaluate the dissipative change in the energy of the heat baths [cf. Eq.~\eqref{eq:exact_stationary_heat_2}]. In fact, this has been criticised as one of the most unsatisfactory features of GKLS-type quantum master equations \cite{PhysRevE.76.031115}. Alternatively, one could think of waiving the secular approximation to work instead with a Redfield master equation \cite{suarez1992memory,gaspard1999slippage}. Although the aforementioned covariances would then cease to be zero, the calculation would continue to yield quantitatively wrong results at very large $ \gamma $---this time simply due to the breakdown of the basic weak-coupling assumption. Ultimately, however, the Redfield approach might improve the GKLS results under moderate residual dissipation \cite{strasberg2016nonequilibrium,strasberg2018fermionic}. Therefore, even in the light of the promising observation made in Sec.~\ref{sec:steady_states} above, great care must still be taken when relying on the RC--GKLS mapping to discuss quantum thermodynamics under non-Markovian dissipation. 

\subsection{Dynamics}\label{sec:dynamics}

\begin{figure}
\includegraphics[width=0.83\linewidth]{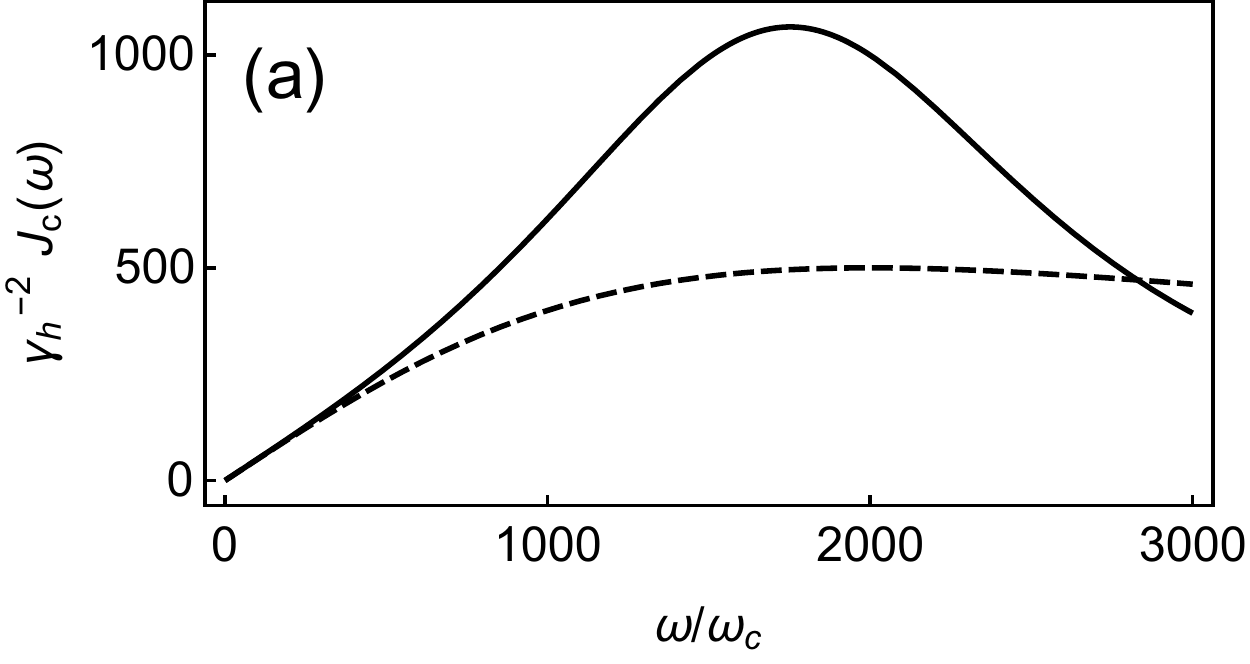}
\includegraphics[width=0.83\linewidth]{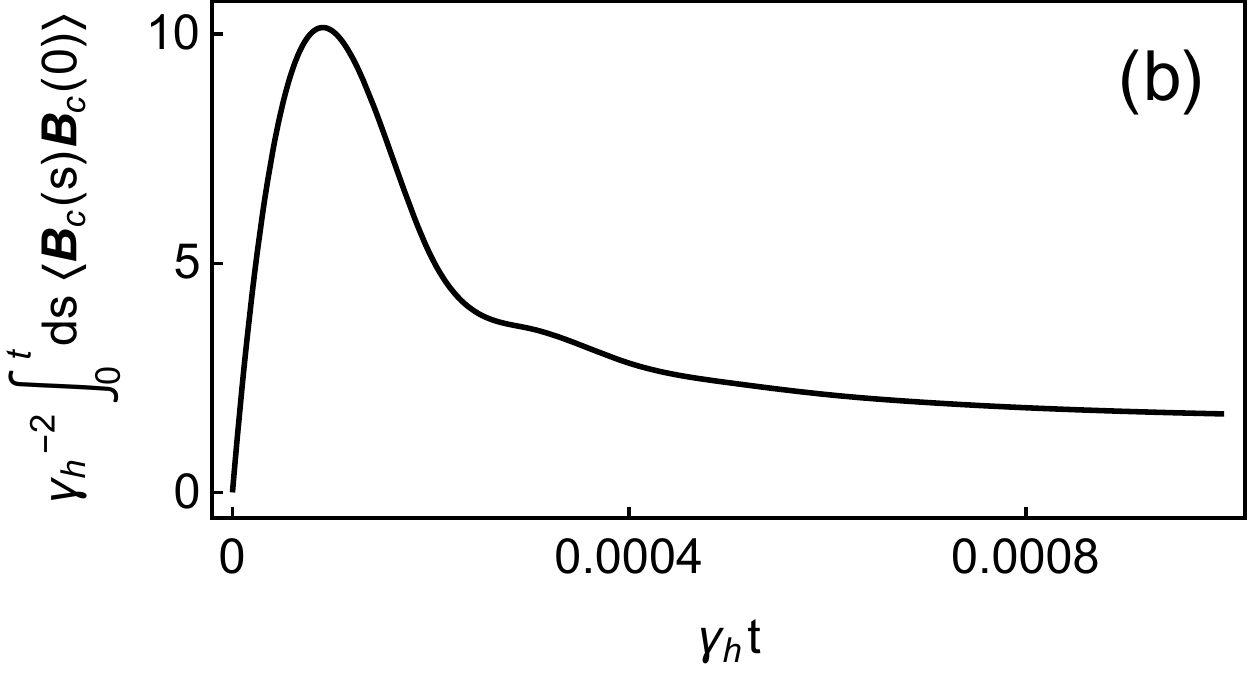}	
\caption{\textbf{(a)} (solid) Structured spectral density $ J_c(\omega) $ with scaled $ \omega_0 $ and $ \lambda $ so as to approximate an overdamped profile in the limit of large $ \gamma $. In particular, $ \lambda^2 = \alpha_1\,\alpha_2\,\gamma $ and $ \omega_0^2 = \alpha_2\,\gamma $, with $ \alpha_1 = \gamma_h = 10^{-3} $, $ \alpha_2 = \Lambda_h = 10^3 $, and $ \gamma = 10^3 $. The limiting Ohmic spectrum (dashed) for $ \gamma\rightarrow\infty $ has been added for comparison. \textbf{(b)} Integrated bath correlation function for the spectral density $ J_c(\omega) $ scaled as in (a). The time elapsed until saturation in the above curve characterises the memory of the cold bath. Hence, if the relevant dynamics occurs over time scales larger than $ \gamma_h t \simeq 10^{-3} $ [cf. Fig.~\hyperref[fig:3]{4(a)} below], we can safely work under the Markov approximation.}\label{fig:2}
\end{figure}

\begin{figure*}[t!]
	\includegraphics[width=0.9\linewidth]{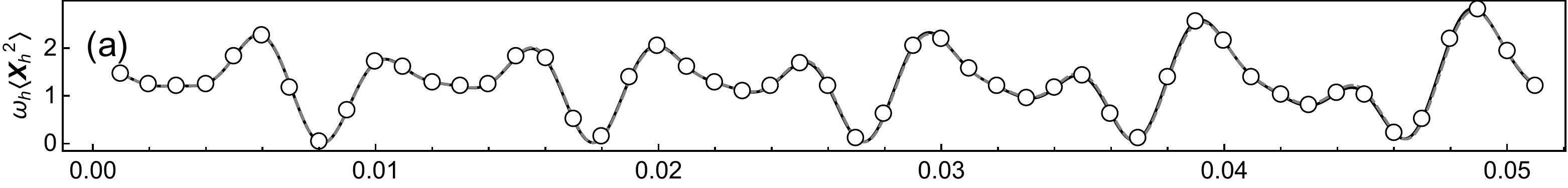}
	\includegraphics[width=0.9\linewidth]{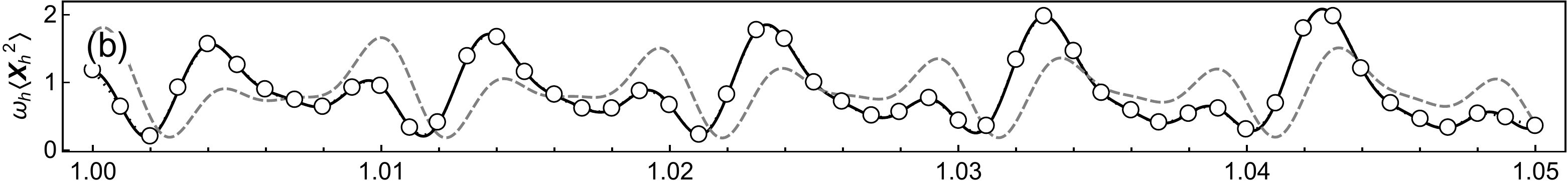}
	\includegraphics[width=0.9\linewidth]{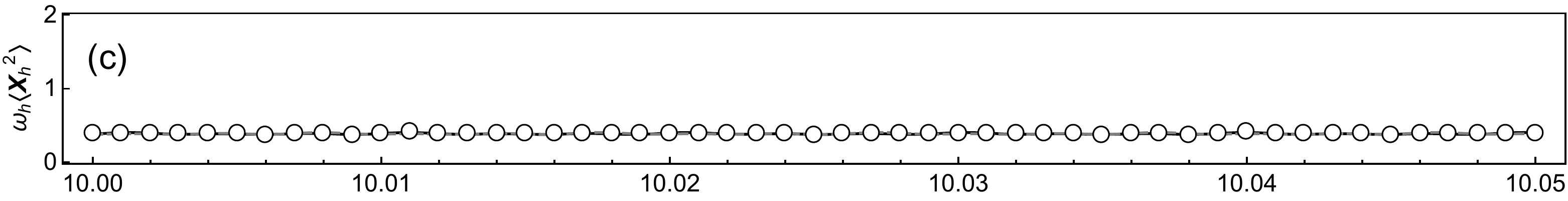}
	\includegraphics[width=0.9\linewidth]{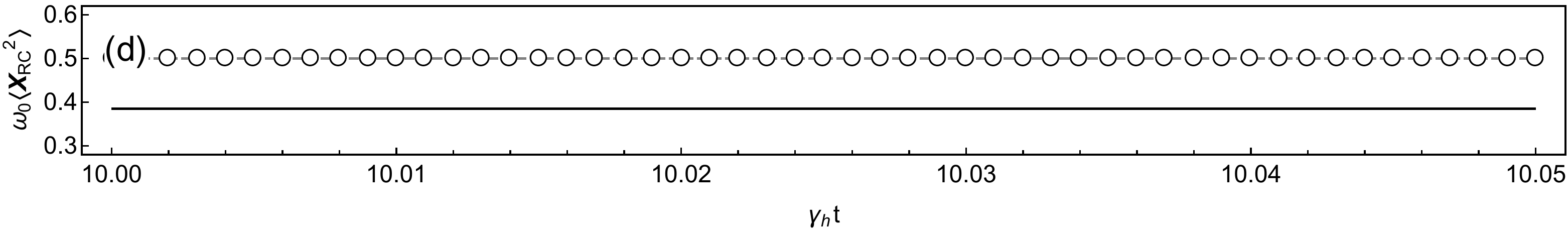}
	\caption{\textbf{(a)--(c)} Time evolution of $ \langle \pmb X_h^2 \rangle $ at different stages of the dynamics, according to a master equation applied directly on the two-node wire (solid black), on the three-node augmented system (open circles), and on an augmented system whose cold frequency has \textit{not} been suitably shifted as per the RC mapping (grey dashed). \textbf{(d)} (solid black) exact steady-state value of $ \langle \pmb X_\text{RC}^2 \rangle$ superimposed to the asymptotic value of this covariance according to the master equation, acting on a shifted (open circles) and unshifted (dashed grey) augmented system. Here, $ \omega_h = 0.1 $, $ \omega_c = 0.5 $, $ k = 0.4 $, $ T_h = 0.6 $, $ T_c = 0.5 $, and the rest of parameters are the same as in Fig.~\ref{fig:2}.} \label{fig:3}
\end{figure*}

One can now ask whether the resilience of the RC--GKLS mapping to strong residual dissipation is exclusively a steady-state feature, or whether it holds throughout the entire dissipative evolution. Unfortunately, we do not have an exact dynamical benchmark---at most, we are able to solve here for the \textit{steady state} of the exact Eq.~\eqref{eq:QLE}. We, therefore, chose parameters so that the original two-node problem \textit{can} be described via a GKLS quantum master equation. {\color{markedgreen} We recall, however, that this type of equation can in principle be solved non-perturbatively at finite times with stochastic propagation techniques \cite{motz2017currents,motz2018rectification}}.  

In particular, we scale $ \omega_0 $ and $ \lambda $ in the structured spectral density $ J_c(\omega) $ in Eq.~\eqref{eq:spectral_density_structured} as $ \lambda^2 = \alpha_1\,\alpha_2\,\gamma $ and $ \omega_0^2 = \gamma\,\alpha_2 $. Taking once again the large friction limit $ \gamma \gg 1 $ leads to the overdamped spectrum $ J_c(\omega) \sim \alpha_1\,\alpha_2\,\omega/(\omega^2 + \alpha_2^2) $ \cite{strasberg2017stochastic}. 

{\color{markedgreen}For our calculations, we will take the numerical values $ \alpha_1 = \gamma_h $ and $ \alpha_2 = \Lambda_h $. Note that $ J_c(\omega) \sim \alpha_1\,\alpha_2\,\omega/(\omega^2 + \alpha_2^2) $  looks like the Ohmic-algebraic $ J_h(\omega) $ introduced above, except for a missing factor $ \alpha_2 $ in the numerator. Hence, while $ \alpha_1 $ takes the numerical value of $ \gamma_h $, it must have units of frequency \textit{squared} instead of frequency. It is $ \alpha_1/\alpha_2 \ll \gamma_h $ which plays the role of the dissipation strength in this case}. In Fig.~\hyperref[fig:2]{3(a)} we plot both the resulting spectral density (solid) along with the Ohmic limiting case of $ \gamma\rightarrow\infty $ (dashed). As it can be seen, for our choice of parameters, the corresponding wire--bath coupling ends up being {\color{markedgreen}at most} $ \mathcal{O}(\gamma_h) \ll 1 $, which would justify the weak-coupling approximation and the use of a perturbative master equation.

The next step towards a GKLS equation is to certify the validity of the Markov approximation: we must ensure that the decay of the bath correlation functions computed in Eq.~\eqref{eq:environmental-correlation} is sufficiently fast when compared to the dynamics of the wire. In Fig.~\hyperref[fig:2]{3(b)} we plot the integrated correlation $ \int_0^t\text{d}s\,\langle\pmb{B}_c(s)\,\pmb{B}_c(0)\rangle $, whose saturation time ($ \gamma_h\tau_c \sim 5 \times 10^{-4} $) is just below the relevant time scale for the dissipative evolution of the wire ($ \gamma_h\tau_w\sim 10^{-3} $) [compare with Fig.~\hyperref[fig:3]{4(a)}]. We thus confidently say that the Markov approximation holds. For the parameters chosen, the secular approximation is also not a problem (cf. caption of Fig.~\ref{fig:3}). Namely, the normal-mode frequencies of the wire are $ \Omega_1 = 0.34 $ and $ \Omega_2 = 0.97 $, while the dissipation rates are both $ \mathcal{O}(\gamma_h) $, which is perturbative. 

We thus take the time evolution of the two-node wire according to the master equation \eqref{eq:GME_general}, as valid approximation to the exact dissipative dynamics, and a good benchmark for the RC mapping. Just like we did in Sec.~\ref{sec:steady-state-and-heat-currents}, we also apply a GKLS master equation to the resulting three-node augmented system; again in spite of the fact that it is totally unjustified (the residual dissipation is $ \gamma = 10^3 $). As pointed out in Sec.~\ref{sec:rc-mapping-nutshell}, initially, we assume no correlations between the reaction coordinate, the wire, and either of the two baths, and initialise the RC in a thermal state at the original temperature of the cold bath. 

Our results are plotted in Fig.~\ref{fig:3}. As we can see, the RC--GKLS mapping (open dots) accurately approximates the dynamics of the covariances of the wire (solid black line), and it does so \textit{during the entire evolution}. However, as expected from the results in Fig.~\hyperref[fig:1]{2(a)}, it fails to capture the covariances of the reaction coordinate itself. We show this in Fig.~\hyperref[fig:3]{4(d)} by comparing the stationary value of $ \langle \pmb X_\text{RC}^2 \rangle $ as predicted by the master equation, with its exact asymptotic value. {\color{markedgreen} It is remarkable, however, that the covariances for the wire are perfectly reproduced in spite of the extremely large friction $ \gamma = 10^3 $. This contrasts with the degradation of fidelity illustrated in Fig.~\ref{fig:1}(a), and is entirely due to our choice of friction-dependent $ \lambda $ and $ \omega_0 $.}

Finally, we also take the opportunity here to illustrate the vital importance of the frequency shift $ \pmb\delta\pmb H_{w\text{--}c} $ on the augmented system (cf. Sec.~\ref{sec:full_hamiltonian}). Note that before the mapping we do not include any shifts in the Hamiltonian of the wire, since we are tackling the original problem via a master equation. However, for the mapping to be an identity, the frequency of the `cold oscillator' must be shifted as in $ \omega_c \mapsto \omega_c^2 + \lambda^2/\omega_0^2 $ when applying the master equation to the augmented system. For our choice of parameters this means tuning it from $ 0.5 $ to $ 0.501 $, which might seem totally negligible. Indeed, the short-time dynamics [cf. Fig.~\hyperref[fig:3]{4(a)}] and the stationary state [cf. Fig.~\hyperref[fig:3]{4(c)}] remain virtually unaffected when the shift is not taken into account (dashed grey lines). At intermediate times, however, the effects of the shift become evident, as shown in Fig.~\hyperref[fig:3]{4(b)}---neglecting it does cause the RC--GKLS mapping to break down. 

\section{Conclusions}\label{sec:conclusions}

We have benchmarked the reaction-coordinate mapping in an exactly solvable linear model, consisting of a two-node chain of harmonic oscillators. These are individually coupled to two baths at different temperatures and thus, support a steady-state heat current. The mapping takes this setup into a three-oscillator augmented system, which is also exactly solvable. The idea, however, is to tackle the augmented system via a weak-coupling Markovian master equation. What we found can be summarised as follows:
\begin{itemize}
	\item The reduction of the stationary state of the augmented system onto the degrees of freedom of the two-node wire---according to the master equation---resembles very closely the exact steady state. This can be so, even in regimes of parameters for which the approximations underpinning the master equation break down; specifically, the secular approximation and even the basic weak-coupling assumption.
	\item Even when the stationary state of the wire is captured faithfully by the master-equation approach, the joint state of all three nodes of the augmented system can differ very substantially from the exact solution of the augmented problem. This happens whenever the underlying approximations cease to be justified. 
	\item More importantly, the non-equilibrium steady state of the wire may be accurately reproduced by the master equation acting on the augmented system and yet, the stationary heat currents obtained from it can be quantitatively and even qualitatively wrong.
	\item At least in the overdamped limit, the reaction-coordinate mapping succeeds in approximating the state of the wire not only asymptotically, but throughout the entire dissipative dynamics.
	
\end{itemize}

In addition, we discussed the subtleties surrounding the frequency renormalisation shifts appearing as a result of the system--environment(s) coupling, and illustrated the importance of using them consistently. We also presented in full detail a consistent Markovian master equation in GKLS form that generalises previous results \cite{gonzalez2017testing}, and can be directly applied to an arbitrary network of of $ N $ harmonic oscillators locally connected to $ M $ heat baths at different temperatures. We explicitly provided the corresponding (Gaussian) non-equilibrium steady state, and the expression for the $ M $ stationary heat currents flowing across the network. {\color{markedgreen} Note that we have focused exclusively on continuous-variable systems in Gaussian states and hence, extending our conclusions to finite-dimensional or non-linear models would require further work.}

Our results have two important consequences when dealing with virtually intractable problems involving nano- and micro-scale systems in non-Markovian baths, such as biological environments. On the one hand, they raise hopes of relying on the combination of `reaction-coordinate mapping' and `weak-coupling master equations' beyond the strict range of applicability of the latter. Although the mapping had been successfully applied to open-systems strongly coupled to highly structured environments \cite{iles2014environmental,iles2016energy,strasberg2016nonequilibrium,newman2017performance,strasberg2018fermionic,wertnik2018optimizing,schaller2018electronic,maguire2018environmental,restrepo2018quantum,restrepo2019electron,mcconnell2019electron,puebla2019spin}, our findings suggest that non-Markovian noise featuring broader power spectra---which so far was though to be out of reach for the mapping---may also be modelled in the exact same manner. On the other hand, however, weak-coupling master equations should \textit{not} be trusted beyond their range of applicability when calculating boundary heat currents---even if these appear to be thermodynamically consistent, they may be serious overestimations. It is pertinent to keep this it in mind when using the reaction-coordinate mapping to extend quantum thermodynamics into the strong coupling regime, an interesting line which currently attracts an increasing attention \cite{strasberg2016nonequilibrium,newman2017performance,restrepo2018quantum,wertnik2018optimizing,strasberg2018fermionic,schaller2018electronic,nazir2018RCmapping}. Put simply, \textit{being able to replicate accurately the exact numerical propagation of an open system with the reaction-coordinate technique does not guarantee that the boundary heat (or particle) currents calculated from the corresponding master equation are equally accurate}. This is our main message. 

{\color{markedgreen}We also note that a closely-related systematic technique has been recently put forward to emulate dissipation into structured environments through GKLS-type master equations \cite{PhysRevLett.120.030402}, which can be used to deal with the strong friction regime. When it comes to extensions of our analysis,} it may be possible to improve on the boundary currents by taking the secular approximation back and working with the full Redfield equation \cite{strasberg2018fermionic,lambert2019virtual}. It would thus be interesting to generalise Eqs.~\eqref{eq:GME_main_covariances}--\eqref{eq:steady_state} and \eqref{eq:steady_heat_GME} to allow for non-secular contributions, and benchmarking those instead. After all, as already mentioned the reaction-coordinate mapping is often combined with Redfield rather than GKLS quantum master equations \cite{iles2014environmental,iles2016energy,strasberg2016nonequilibrium,strasberg2018fermionic,lambert2019virtual}. It is important to bear in mind, however, that Redfield equations may not only violate complete positivity, but even \textit{positivity} alone \cite{suarez1992memory,gaspard1999slippage}, which seriously compromises the consistency of any quantum-thermodynamic variables derived from it. This generalisation lies, however, beyond the scope of this paper, and will be tackled elsewhere.

\section*{Acknowledgements}

We gratefully acknowledge Janet Anders, James Cresser, Ronnie Kosloff, Neill Lambert, Ahsan Nazir, Philipp Strasberg and Tommaso Tufarelli for useful discussions. This project was funded by the European Research Council (StG GQCOP, Grant No. 637352), the London Mathematical Society (Scheme 3 Grant No. 31826), and the US National Science Foundation (Grant No. NSF PHY1748958). LAC and BM thank the Kavli Institute for Theoretical Physics for their warm hospitality during the program ``Thermodynamics of quantum systems: Measurement, engines, and control'' and the associated conference.

\section*{References}
%\bibliographystyle{aipnum4-1}
%\bibliography{pushing}

\begin{thebibliography}{84}%
	\makeatletter
	\providecommand \@ifxundefined [1]{%
		\@ifx{#1\undefined}
	}%
	\providecommand \@ifnum [1]{%
		\ifnum #1\expandafter \@firstoftwo
		\else \expandafter \@secondoftwo
		\fi
	}%
	\providecommand \@ifx [1]{%
		\ifx #1\expandafter \@firstoftwo
		\else \expandafter \@secondoftwo
		\fi
	}%
	\providecommand \natexlab [1]{#1}%
	\providecommand \enquote  [1]{``#1''}%
	\providecommand \bibnamefont  [1]{#1}%
	\providecommand \bibfnamefont [1]{#1}%
	\providecommand \citenamefont [1]{#1}%
	\providecommand \href@noop [0]{\@secondoftwo}%
	\providecommand \href [0]{\begingroup \@sanitize@url \@href}%
	\providecommand \@href[1]{\@@startlink{#1}\@@href}%
	\providecommand \@@href[1]{\endgroup#1\@@endlink}%
	\providecommand \@sanitize@url [0]{\catcode `\\12\catcode `\$12\catcode
		`\&12\catcode `\#12\catcode `\^12\catcode `\_12\catcode `\%12\relax}%
	\providecommand \@@startlink[1]{}%
	\providecommand \@@endlink[0]{}%
	\providecommand \url  [0]{\begingroup\@sanitize@url \@url }%
	\providecommand \@url [1]{\endgroup\@href {#1}{\urlprefix }}%
	\providecommand \urlprefix  [0]{URL }%
	\providecommand \Eprint [0]{\href }%
	\providecommand \doibase [0]{http://dx.doi.org/}%
	\providecommand \selectlanguage [0]{\@gobble}%
	\providecommand \bibinfo  [0]{\@secondoftwo}%
	\providecommand \bibfield  [0]{\@secondoftwo}%
	\providecommand \translation [1]{[#1]}%
	\providecommand \BibitemOpen [0]{}%
	\providecommand \bibitemStop [0]{}%
	\providecommand \bibitemNoStop [0]{.\EOS\space}%
	\providecommand \EOS [0]{\spacefactor3000\relax}%
	\providecommand \BibitemShut  [1]{\csname bibitem#1\endcsname}%
	\let\auto@bib@innerbib\@empty
	%</preamble>
	\bibitem [{\citenamefont {Garg}, \citenamefont {Onuchic},\ and\ \citenamefont
		{Ambegaokar}(1985)}]{garg1985effect}%
	\BibitemOpen
	\bibfield  {author} {\bibinfo {author} {\bibfnamefont {A.}~\bibnamefont
			{Garg}}, \bibinfo {author} {\bibfnamefont {J.~N.}\ \bibnamefont {Onuchic}}, \
		and\ \bibinfo {author} {\bibfnamefont {V.}~\bibnamefont {Ambegaokar}},\
	}\href {\doibase 10.1063/1.449017} {\bibfield  {journal} {\bibinfo  {journal}
			{J. Chem. Phys.}\ }\textbf {\bibinfo {volume} {83}},\ \bibinfo {pages} {4491}
		(\bibinfo {year} {1985})}\BibitemShut {NoStop}%
	\bibitem [{\citenamefont {Hartmann}, \citenamefont {Goychuk},\ and\
		\citenamefont {H{\"a}nggi}(2000)}]{hartmann2000controlling}%
	\BibitemOpen
	\bibfield  {author} {\bibinfo {author} {\bibfnamefont {L.}~\bibnamefont
			{Hartmann}}, \bibinfo {author} {\bibfnamefont {I.}~\bibnamefont {Goychuk}}, \
		and\ \bibinfo {author} {\bibfnamefont {P.}~\bibnamefont {H{\"a}nggi}},\
	}\href {\doibase 10.1063/1.1326049} {\bibfield  {journal} {\bibinfo
			{journal} {J. Chem. Phys.}\ }\textbf {\bibinfo {volume} {113}},\ \bibinfo
		{pages} {11159} (\bibinfo {year} {2000})}\BibitemShut {NoStop}%
	\bibitem [{\citenamefont {Roden}\ \emph {et~al.}(2012)\citenamefont {Roden},
		\citenamefont {Strunz}, \citenamefont {Whaley},\ and\ \citenamefont
		{Eisfeld}}]{roden2012accounting}%
	\BibitemOpen
	\bibfield  {author} {\bibinfo {author} {\bibfnamefont {J.}~\bibnamefont
			{Roden}}, \bibinfo {author} {\bibfnamefont {W.~T.}\ \bibnamefont {Strunz}},
		\bibinfo {author} {\bibfnamefont {K.~B.}\ \bibnamefont {Whaley}}, \ and\
		\bibinfo {author} {\bibfnamefont {A.}~\bibnamefont {Eisfeld}},\ }\href
	{\doibase 10.1063/1.4765329} {\bibfield  {journal} {\bibinfo  {journal} {J.
				Chem. Phys.}\ }\textbf {\bibinfo {volume} {137}},\ \bibinfo {pages} {204110}
		(\bibinfo {year} {2012})}\BibitemShut {NoStop}%
	\bibitem [{\citenamefont {Binder}\ \emph {et~al.}(2018)\citenamefont {Binder},
		\citenamefont {Correa}, \citenamefont {Gogolin}, \citenamefont {Anders},\
		and\ \citenamefont {Adesso}}]{binder2018thermodynamics}%
	\BibitemOpen
	\bibinfo {editor} {\bibfnamefont {F.}~\bibnamefont {Binder}}, \bibinfo
	{editor} {\bibfnamefont {L.~A.}\ \bibnamefont {Correa}}, \bibinfo {editor}
	{\bibfnamefont {C.}~\bibnamefont {Gogolin}}, \bibinfo {editor} {\bibfnamefont
		{J.}~\bibnamefont {Anders}}, \ and\ \bibinfo {editor} {\bibfnamefont
		{G.}~\bibnamefont {Adesso}},\ eds.,\ \href@noop {} {\emph {\bibinfo {title}
			{Thermodynamics in the quantum regime}}},\ Fundamental Theories of Physics\
	(\bibinfo  {publisher} {Springer},\ \bibinfo {year} {2018})\BibitemShut
	{NoStop}%
	\bibitem [{\citenamefont {Alicki}\ \emph {et~al.}(2004)\citenamefont {Alicki},
		\citenamefont {Horodecki}, \citenamefont {Horodecki}, \citenamefont
		{Horodecki}, \citenamefont {Jacak},\ and\ \citenamefont
		{Machnikowski}}]{alicki2004optimal}%
	\BibitemOpen
	\bibfield  {author} {\bibinfo {author} {\bibfnamefont {R.}~\bibnamefont
			{Alicki}}, \bibinfo {author} {\bibfnamefont {M.}~\bibnamefont {Horodecki}},
		\bibinfo {author} {\bibfnamefont {P.}~\bibnamefont {Horodecki}}, \bibinfo
		{author} {\bibfnamefont {R.}~\bibnamefont {Horodecki}}, \bibinfo {author}
		{\bibfnamefont {L.}~\bibnamefont {Jacak}}, \ and\ \bibinfo {author}
		{\bibfnamefont {P.}~\bibnamefont {Machnikowski}},\ }\href {\doibase
		10.1103/PhysRevA.70.010501} {\bibfield  {journal} {\bibinfo  {journal} {Phys.
				Rev. A}\ }\textbf {\bibinfo {volume} {70}},\ \bibinfo {pages} {010501}
		(\bibinfo {year} {2004})}\BibitemShut {NoStop}%
	\bibitem [{\citenamefont {McCutcheon}\ and\ \citenamefont
		{Nazir}(2010)}]{mccutcheon2010quantum}%
	\BibitemOpen
	\bibfield  {author} {\bibinfo {author} {\bibfnamefont {D.~P.}\ \bibnamefont
			{McCutcheon}}\ and\ \bibinfo {author} {\bibfnamefont {A.}~\bibnamefont
			{Nazir}},\ }\href {\doibase 10.1088/1367-2630/12/11/113042} {\bibfield
		{journal} {\bibinfo  {journal} {New J. Phys.}\ }\textbf {\bibinfo {volume}
			{12}},\ \bibinfo {pages} {113042} (\bibinfo {year} {2010})}\BibitemShut
	{NoStop}%
	\bibitem [{\citenamefont {Higgins}, \citenamefont {Lovett},\ and\ \citenamefont
		{Gauger}(2013)}]{higgins2013thermometry}%
	\BibitemOpen
	\bibfield  {author} {\bibinfo {author} {\bibfnamefont {K.~D.~B.}\
			\bibnamefont {Higgins}}, \bibinfo {author} {\bibfnamefont {B.~W.}\
			\bibnamefont {Lovett}}, \ and\ \bibinfo {author} {\bibfnamefont {E.~M.}\
			\bibnamefont {Gauger}},\ }\href {\doibase 10.1103/PhysRevB.88.155409}
	{\bibfield  {journal} {\bibinfo  {journal} {Phys. Rev. B}\ }\textbf {\bibinfo
			{volume} {88}},\ \bibinfo {pages} {155409} (\bibinfo {year}
		{2013})}\BibitemShut {NoStop}%
	\bibitem [{\citenamefont {de~Vega}\ and\ \citenamefont
		{Alonso}(2017)}]{devega2017nonmarkovian}%
	\BibitemOpen
	\bibfield  {author} {\bibinfo {author} {\bibfnamefont {I.}~\bibnamefont
			{de~Vega}}\ and\ \bibinfo {author} {\bibfnamefont {D.}~\bibnamefont
			{Alonso}},\ }\href {\doibase 10.1103/RevModPhys.89.015001} {\bibfield
		{journal} {\bibinfo  {journal} {Rev. Mod. Phys.}\ }\textbf {\bibinfo {volume}
			{89}},\ \bibinfo {pages} {015001} (\bibinfo {year} {2017})}\BibitemShut
	{NoStop}%
	\bibitem [{\citenamefont {Feynman}\ and\ \citenamefont
		{Vernon~Jr}(2000)}]{feynman2000theory}%
	\BibitemOpen
	\bibfield  {author} {\bibinfo {author} {\bibfnamefont {R.~P.}\ \bibnamefont
			{Feynman}}\ and\ \bibinfo {author} {\bibfnamefont {F.}~\bibnamefont
			{Vernon~Jr}},\ }\href {\doibase 10.1006/aphy.2000.6017} {\bibfield  {journal}
		{\bibinfo  {journal} {Ann. Phys. (N. Y.)}\ }\textbf {\bibinfo {volume}
			{281}},\ \bibinfo {pages} {547} (\bibinfo {year} {2000})}\BibitemShut
	{NoStop}%
	\bibitem [{\citenamefont {Hu}, \citenamefont {Paz},\ and\ \citenamefont
		{Zhang}(1992)}]{hu1992brownian}%
	\BibitemOpen
	\bibfield  {author} {\bibinfo {author} {\bibfnamefont {B.~L.}\ \bibnamefont
			{Hu}}, \bibinfo {author} {\bibfnamefont {J.~P.}\ \bibnamefont {Paz}}, \ and\
		\bibinfo {author} {\bibfnamefont {Y.}~\bibnamefont {Zhang}},\ }\href
	{\doibase 10.1103/PhysRevD.45.2843} {\bibfield  {journal} {\bibinfo
			{journal} {Phys. Rev. D}\ }\textbf {\bibinfo {volume} {45}},\ \bibinfo
		{pages} {2843} (\bibinfo {year} {1992})}\BibitemShut {NoStop}%
	\bibitem [{\citenamefont {Tanimura}(1990)}]{tanimura1990heom}%
	\BibitemOpen
	\bibfield  {author} {\bibinfo {author} {\bibfnamefont {Y.}~\bibnamefont
			{Tanimura}},\ }\href {\doibase 10.1103/PhysRevA.41.6676} {\bibfield
		{journal} {\bibinfo  {journal} {Phys. Rev. A}\ }\textbf {\bibinfo {volume}
			{41}},\ \bibinfo {pages} {6676} (\bibinfo {year} {1990})}\BibitemShut
	{NoStop}%
	\bibitem [{\citenamefont {Stockburger}\ and\ \citenamefont
		{Grabert}(2002)}]{stockburger2002nonmarkov}%
	\BibitemOpen
	\bibfield  {author} {\bibinfo {author} {\bibfnamefont {J.~T.}\ \bibnamefont
			{Stockburger}}\ and\ \bibinfo {author} {\bibfnamefont {H.}~\bibnamefont
			{Grabert}},\ }\href {\doibase 10.1103/PhysRevLett.88.170407} {\bibfield
		{journal} {\bibinfo  {journal} {Phys. Rev. Lett.}\ }\textbf {\bibinfo
			{volume} {88}},\ \bibinfo {pages} {170407} (\bibinfo {year}
		{2002})}\BibitemShut {NoStop}%
	\bibitem [{\citenamefont {Alonso}\ and\ \citenamefont
		{de~Vega}(2005)}]{devega2005multiple}%
	\BibitemOpen
	\bibfield  {author} {\bibinfo {author} {\bibfnamefont {D.}~\bibnamefont
			{Alonso}}\ and\ \bibinfo {author} {\bibfnamefont {I.}~\bibnamefont
			{de~Vega}},\ }\href {\doibase 10.1103/PhysRevLett.94.200403} {\bibfield
		{journal} {\bibinfo  {journal} {Phys. Rev. Lett.}\ }\textbf {\bibinfo
			{volume} {94}},\ \bibinfo {pages} {200403} (\bibinfo {year}
		{2005})}\BibitemShut {NoStop}%
	\bibitem [{\citenamefont {Wagner}(1986)}]{wagner1986unitary}%
	\BibitemOpen
	\bibfield  {author} {\bibinfo {author} {\bibfnamefont {M.}~\bibnamefont
			{Wagner}},\ }\href@noop {} {\emph {\bibinfo {title} {Unitary transformations
				in solid state physics}}},\ \bibinfo {series} {Modern Problems in Condensed
		Matter Sciences}, Vol.~\bibinfo {volume} {15}\ (\bibinfo  {publisher}
	{North-Holland},\ \bibinfo {year} {1986})\BibitemShut {NoStop}%
	\bibitem [{\citenamefont {W\"urger}(1998)}]{wurger1998polaron}%
	\BibitemOpen
	\bibfield  {author} {\bibinfo {author} {\bibfnamefont {A.}~\bibnamefont
			{W\"urger}},\ }\href {\doibase 10.1103/PhysRevB.57.347} {\bibfield  {journal}
		{\bibinfo  {journal} {Phys. Rev. B}\ }\textbf {\bibinfo {volume} {57}},\
		\bibinfo {pages} {347} (\bibinfo {year} {1998})}\BibitemShut {NoStop}%
	\bibitem [{\citenamefont {Garraway}(1997)}]{garraway1997pseudomodes}%
	\BibitemOpen
	\bibfield  {author} {\bibinfo {author} {\bibfnamefont {B.~M.}\ \bibnamefont
			{Garraway}},\ }\href {\doibase 10.1103/PhysRevA.55.4636} {\bibfield
		{journal} {\bibinfo  {journal} {Phys. Rev. A}\ }\textbf {\bibinfo {volume}
			{55}},\ \bibinfo {pages} {4636} (\bibinfo {year} {1997})}\BibitemShut
	{NoStop}%
	\bibitem [{\citenamefont {Martinazzo}\ \emph {et~al.}(2011)\citenamefont
		{Martinazzo}, \citenamefont {Vacchini}, \citenamefont {Hughes},\ and\
		\citenamefont {Burghardt}}]{martinazzo2011universal}%
	\BibitemOpen
	\bibfield  {author} {\bibinfo {author} {\bibfnamefont {R.}~\bibnamefont
			{Martinazzo}}, \bibinfo {author} {\bibfnamefont {B.}~\bibnamefont
			{Vacchini}}, \bibinfo {author} {\bibfnamefont {K.}~\bibnamefont {Hughes}}, \
		and\ \bibinfo {author} {\bibfnamefont {I.}~\bibnamefont {Burghardt}},\ }\href
	{\doibase 10.1063/1.3532408} {\bibfield  {journal} {\bibinfo  {journal} {J.
				Chem. Phys.}\ }\textbf {\bibinfo {volume} {134}},\ \bibinfo {pages} {011101}
		(\bibinfo {year} {2011})}\BibitemShut {NoStop}%
	\bibitem [{\citenamefont {Woods}\ \emph {et~al.}(2014)\citenamefont {Woods},
		\citenamefont {Groux}, \citenamefont {Chin}, \citenamefont {Huelga},\ and\
		\citenamefont {Plenio}}]{woods2014mappings}%
	\BibitemOpen
	\bibfield  {author} {\bibinfo {author} {\bibfnamefont {M.}~\bibnamefont
			{Woods}}, \bibinfo {author} {\bibfnamefont {R.}~\bibnamefont {Groux}},
		\bibinfo {author} {\bibfnamefont {A.}~\bibnamefont {Chin}}, \bibinfo {author}
		{\bibfnamefont {S.}~\bibnamefont {Huelga}}, \ and\ \bibinfo {author}
		{\bibfnamefont {M.~B.}\ \bibnamefont {Plenio}},\ }\href {\doibase
		10.1063/1.4866769} {\bibfield  {journal} {\bibinfo  {journal} {J. Math.
				Phys.}\ }\textbf {\bibinfo {volume} {55}},\ \bibinfo {pages} {032101}
		(\bibinfo {year} {2014})}\BibitemShut {NoStop}%
	\bibitem [{\citenamefont {Iles-Smith}, \citenamefont {Lambert},\ and\
		\citenamefont {Nazir}(2014)}]{iles2014environmental}%
	\BibitemOpen
	\bibfield  {author} {\bibinfo {author} {\bibfnamefont {J.}~\bibnamefont
			{Iles-Smith}}, \bibinfo {author} {\bibfnamefont {N.}~\bibnamefont {Lambert}},
		\ and\ \bibinfo {author} {\bibfnamefont {A.}~\bibnamefont {Nazir}},\ }\href
	{\doibase 10.1103/PhysRevA.90.032114} {\bibfield  {journal} {\bibinfo
			{journal} {Phys. Rev. A}\ }\textbf {\bibinfo {volume} {90}},\ \bibinfo
		{pages} {032114} (\bibinfo {year} {2014})}\BibitemShut {NoStop}%
	\bibitem [{\citenamefont {Nazir}\ and\ \citenamefont
		{Schaller}(2018)}]{nazir2018RCmapping}%
	\BibitemOpen
	\bibfield  {author} {\bibinfo {author} {\bibfnamefont {A.}~\bibnamefont
			{Nazir}}\ and\ \bibinfo {author} {\bibfnamefont {G.}~\bibnamefont
			{Schaller}},\ }\enquote {\bibinfo {title} {Thermodynamics in the quantum
			regime},}\ \ (\bibinfo  {publisher} {Springer},\ \bibinfo {year} {2018})\
	Chap.~\bibinfo {chapter} {23}, p.\ \bibinfo {pages} {551}\BibitemShut
	{NoStop}%
	\bibitem [{\citenamefont {Hughes}, \citenamefont {Christ},\ and\ \citenamefont
		{Burghardt}(2009)}]{hughes2009effective}%
	\BibitemOpen
	\bibfield  {author} {\bibinfo {author} {\bibfnamefont {K.~H.}\ \bibnamefont
			{Hughes}}, \bibinfo {author} {\bibfnamefont {C.~D.}\ \bibnamefont {Christ}},
		\ and\ \bibinfo {author} {\bibfnamefont {I.}~\bibnamefont {Burghardt}},\
	}\href {\doibase 10.1063/1.3159671} {\bibfield  {journal} {\bibinfo
			{journal} {J. Chem. Phys.}\ }\textbf {\bibinfo {volume} {131}},\ \bibinfo
		{pages} {024109} (\bibinfo {year} {2009})}\BibitemShut {NoStop}%
	\bibitem [{Note1()}]{Note1}%
	\BibitemOpen
	\bibinfo {note} {A Fokker--Plank equation may be derived in the opposite
		large-friction limit \cite
		{garg1985effect,iles2014environmental}.}\BibitemShut {Stop}%
	\bibitem [{\citenamefont {Iles-Smith}\ \emph {et~al.}(2016)\citenamefont
		{Iles-Smith}, \citenamefont {Dijkstra}, \citenamefont {Lambert},\ and\
		\citenamefont {Nazir}}]{iles2016energy}%
	\BibitemOpen
	\bibfield  {author} {\bibinfo {author} {\bibfnamefont {J.}~\bibnamefont
			{Iles-Smith}}, \bibinfo {author} {\bibfnamefont {A.~G.}\ \bibnamefont
			{Dijkstra}}, \bibinfo {author} {\bibfnamefont {N.}~\bibnamefont {Lambert}}, \
		and\ \bibinfo {author} {\bibfnamefont {A.}~\bibnamefont {Nazir}},\ }\href
	{\doibase 10.1063/1.4940218} {\bibfield  {journal} {\bibinfo  {journal} {J.
				Chem. Phys.}\ }\textbf {\bibinfo {volume} {144}},\ \bibinfo {pages} {044110}
		(\bibinfo {year} {2016})}\BibitemShut {NoStop}%
	\bibitem [{\citenamefont {Strasberg}\ \emph {et~al.}(2016)\citenamefont
		{Strasberg}, \citenamefont {Schaller}, \citenamefont {Lambert},\ and\
		\citenamefont {Brandes}}]{strasberg2016nonequilibrium}%
	\BibitemOpen
	\bibfield  {author} {\bibinfo {author} {\bibfnamefont {P.}~\bibnamefont
			{Strasberg}}, \bibinfo {author} {\bibfnamefont {G.}~\bibnamefont {Schaller}},
		\bibinfo {author} {\bibfnamefont {N.}~\bibnamefont {Lambert}}, \ and\
		\bibinfo {author} {\bibfnamefont {T.}~\bibnamefont {Brandes}},\ }\href
	{\doibase 10.1088/1367-2630/18/7/073007} {\bibfield  {journal} {\bibinfo
			{journal} {New J. Phys.}\ }\textbf {\bibinfo {volume} {18}},\ \bibinfo
		{pages} {073007} (\bibinfo {year} {2016})}\BibitemShut {NoStop}%
	\bibitem [{\citenamefont {Restrepo}\ \emph {et~al.}(2018)\citenamefont
		{Restrepo}, \citenamefont {Cerrillo}, \citenamefont {Strasberg},\ and\
		\citenamefont {Schaller}}]{restrepo2018quantum}%
	\BibitemOpen
	\bibfield  {author} {\bibinfo {author} {\bibfnamefont {S.}~\bibnamefont
			{Restrepo}}, \bibinfo {author} {\bibfnamefont {J.}~\bibnamefont {Cerrillo}},
		\bibinfo {author} {\bibfnamefont {P.}~\bibnamefont {Strasberg}}, \ and\
		\bibinfo {author} {\bibfnamefont {G.}~\bibnamefont {Schaller}},\ }\href
	{\doibase 10.1088/1367-2630/aac583} {\bibfield  {journal} {\bibinfo
			{journal} {New J. Phys.}\ }\textbf {\bibinfo {volume} {20}},\ \bibinfo
		{pages} {053063} (\bibinfo {year} {2018})}\BibitemShut {NoStop}%
	\bibitem [{\citenamefont {Wertnik}\ \emph {et~al.}(2018)\citenamefont
		{Wertnik}, \citenamefont {Chin}, \citenamefont {Nori},\ and\ \citenamefont
		{Lambert}}]{wertnik2018optimizing}%
	\BibitemOpen
	\bibfield  {author} {\bibinfo {author} {\bibfnamefont {M.}~\bibnamefont
			{Wertnik}}, \bibinfo {author} {\bibfnamefont {A.}~\bibnamefont {Chin}},
		\bibinfo {author} {\bibfnamefont {F.}~\bibnamefont {Nori}}, \ and\ \bibinfo
		{author} {\bibfnamefont {N.}~\bibnamefont {Lambert}},\ }\href {\doibase
		10.1063/1.5040898} {\bibfield  {journal} {\bibinfo  {journal} {J. Chem.
				Phys.}\ }\textbf {\bibinfo {volume} {149}},\ \bibinfo {pages} {084112}
		(\bibinfo {year} {2018})}\BibitemShut {NoStop}%
	\bibitem [{\citenamefont {Maguire}, \citenamefont {Iles-Smith},\ and\
		\citenamefont {Nazir}(2018)}]{maguire2018environmental}%
	\BibitemOpen
	\bibfield  {author} {\bibinfo {author} {\bibfnamefont {H.}~\bibnamefont
			{Maguire}}, \bibinfo {author} {\bibfnamefont {J.}~\bibnamefont {Iles-Smith}},
		\ and\ \bibinfo {author} {\bibfnamefont {A.}~\bibnamefont {Nazir}},\ }\href
	{https://arxiv.org/abs/1812.04502} {\bibfield  {journal} {\bibinfo  {journal}
			{arXiv preprint arXiv:1812.04502}\ } (\bibinfo {year} {2018})}\BibitemShut
	{NoStop}%
	\bibitem [{\citenamefont {Puebla}\ \emph {et~al.}(2019)\citenamefont {Puebla},
		\citenamefont {Zicari}, \citenamefont {Arrazola}, \citenamefont {Solano},
		\citenamefont {Paternostro},\ and\ \citenamefont
		{Casanova}}]{puebla2019spin}%
	\BibitemOpen
	\bibfield  {author} {\bibinfo {author} {\bibfnamefont {R.}~\bibnamefont
			{Puebla}}, \bibinfo {author} {\bibfnamefont {G.}~\bibnamefont {Zicari}},
		\bibinfo {author} {\bibfnamefont {I.}~\bibnamefont {Arrazola}}, \bibinfo
		{author} {\bibfnamefont {E.}~\bibnamefont {Solano}}, \bibinfo {author}
		{\bibfnamefont {M.}~\bibnamefont {Paternostro}}, \ and\ \bibinfo {author}
		{\bibfnamefont {J.}~\bibnamefont {Casanova}},\ }\href {\doibase
		10.3390/sym11050695} {\bibfield  {journal} {\bibinfo  {journal} {Symmetry}\
		}\textbf {\bibinfo {volume} {11}},\ \bibinfo {pages} {695} (\bibinfo {year}
		{2019})}\BibitemShut {NoStop}%
	\bibitem [{\citenamefont {Martensen}\ and\ \citenamefont
		{Schaller}(2019)}]{martensen2019transmission}%
	\BibitemOpen
	\bibfield  {author} {\bibinfo {author} {\bibfnamefont {N.}~\bibnamefont
			{Martensen}}\ and\ \bibinfo {author} {\bibfnamefont {G.}~\bibnamefont
			{Schaller}},\ }\href {\doibase 10.1140/epjb/e2019-90585-0} {\bibfield
		{journal} {\bibinfo  {journal} {Eur. Phys. J. B}\ }\textbf {\bibinfo {volume}
			{92}},\ \bibinfo {pages} {30} (\bibinfo {year} {2019})}\BibitemShut {NoStop}%
	\bibitem [{\citenamefont {McConnell}\ and\ \citenamefont
		{Nazir}(2019)}]{mcconnell2019electron}%
	\BibitemOpen
	\bibfield  {author} {\bibinfo {author} {\bibfnamefont {C.}~\bibnamefont
			{McConnell}}\ and\ \bibinfo {author} {\bibfnamefont {A.}~\bibnamefont
			{Nazir}},\ }\href {https://arxiv.org/abs/1903.05264} {\bibfield  {journal}
		{\bibinfo  {journal} {arXiv preprint arXiv:1903.05264}\ } (\bibinfo {year}
		{2019})}\BibitemShut {NoStop}%
	\bibitem [{\citenamefont {Lambert}\ \emph {et~al.}(2019)\citenamefont
		{Lambert}, \citenamefont {Ahmed}, \citenamefont {Cirio},\ and\ \citenamefont
		{Nori}}]{lambert2019virtual}%
	\BibitemOpen
	\bibfield  {author} {\bibinfo {author} {\bibfnamefont {N.}~\bibnamefont
			{Lambert}}, \bibinfo {author} {\bibfnamefont {S.}~\bibnamefont {Ahmed}},
		\bibinfo {author} {\bibfnamefont {M.}~\bibnamefont {Cirio}}, \ and\ \bibinfo
		{author} {\bibfnamefont {F.}~\bibnamefont {Nori}},\ }\href
	{https://arxiv.org/abs/1903.05892} {\bibfield  {journal} {\bibinfo  {journal}
			{arXiv preprint arXiv:1903.05892}\ } (\bibinfo {year} {2019})}\BibitemShut
	{NoStop}%
	\bibitem [{\citenamefont {Strasberg}\ \emph {et~al.}(2018)\citenamefont
		{Strasberg}, \citenamefont {Schaller}, \citenamefont {Schmidt},\ and\
		\citenamefont {Esposito}}]{strasberg2018fermionic}%
	\BibitemOpen
	\bibfield  {author} {\bibinfo {author} {\bibfnamefont {P.}~\bibnamefont
			{Strasberg}}, \bibinfo {author} {\bibfnamefont {G.}~\bibnamefont {Schaller}},
		\bibinfo {author} {\bibfnamefont {T.~L.}\ \bibnamefont {Schmidt}}, \ and\
		\bibinfo {author} {\bibfnamefont {M.}~\bibnamefont {Esposito}},\ }\href
	{\doibase 10.1103/PhysRevB.97.205405} {\bibfield  {journal} {\bibinfo
			{journal} {Phys. Rev. B}\ }\textbf {\bibinfo {volume} {97}},\ \bibinfo
		{pages} {205405} (\bibinfo {year} {2018})}\BibitemShut {NoStop}%
	\bibitem [{\citenamefont {Schaller}\ \emph {et~al.}(2018)\citenamefont
		{Schaller}, \citenamefont {Cerrillo}, \citenamefont {Engelhardt},\ and\
		\citenamefont {Strasberg}}]{schaller2018electronic}%
	\BibitemOpen
	\bibfield  {author} {\bibinfo {author} {\bibfnamefont {G.}~\bibnamefont
			{Schaller}}, \bibinfo {author} {\bibfnamefont {J.}~\bibnamefont {Cerrillo}},
		\bibinfo {author} {\bibfnamefont {G.}~\bibnamefont {Engelhardt}}, \ and\
		\bibinfo {author} {\bibfnamefont {P.}~\bibnamefont {Strasberg}},\ }\href
	{\doibase 10.1103/PhysRevB.97.195104} {\bibfield  {journal} {\bibinfo
			{journal} {Phys. Rev. B}\ }\textbf {\bibinfo {volume} {97}},\ \bibinfo
		{pages} {195104} (\bibinfo {year} {2018})}\BibitemShut {NoStop}%
	\bibitem [{\citenamefont {Restrepo}\ \emph {et~al.}(2019)\citenamefont
		{Restrepo}, \citenamefont {B{\"o}hling}, \citenamefont {Cerrillo},\ and\
		\citenamefont {Schaller}}]{restrepo2019electron}%
	\BibitemOpen
	\bibfield  {author} {\bibinfo {author} {\bibfnamefont {S.}~\bibnamefont
			{Restrepo}}, \bibinfo {author} {\bibfnamefont {S.}~\bibnamefont
			{B{\"o}hling}}, \bibinfo {author} {\bibfnamefont {J.}~\bibnamefont
			{Cerrillo}}, \ and\ \bibinfo {author} {\bibfnamefont {G.}~\bibnamefont
			{Schaller}},\ }\href {https://arxiv.org/abs/1905.00581} {\bibfield  {journal}
		{\bibinfo  {journal} {arXiv preprint arXiv:1905.00581}\ } (\bibinfo {year}
		{2019})}\BibitemShut {NoStop}%
	\bibitem [{Note2()}]{Note2}%
	\BibitemOpen
	\bibinfo {note} {Recently, the RC mapping has been shown to break down at
		large friction in the zero-temperature limit \cite {lambert2019virtual}. Our
		calculations here are, however, limited to \protect \textit {finite}
		temperatures.}\BibitemShut {Stop}%
	\bibitem [{\citenamefont {Gonz{\'a}lez}\ \emph {et~al.}(2017)\citenamefont
		{Gonz{\'a}lez}, \citenamefont {Correa}, \citenamefont {Nocerino},
		\citenamefont {Palao}, \citenamefont {Alonso},\ and\ \citenamefont
		{Adesso}}]{gonzalez2017testing}%
	\BibitemOpen
	\bibfield  {author} {\bibinfo {author} {\bibfnamefont {J.~O.}\ \bibnamefont
			{Gonz{\'a}lez}}, \bibinfo {author} {\bibfnamefont {L.~A.}\ \bibnamefont
			{Correa}}, \bibinfo {author} {\bibfnamefont {G.}~\bibnamefont {Nocerino}},
		\bibinfo {author} {\bibfnamefont {J.~P.}\ \bibnamefont {Palao}}, \bibinfo
		{author} {\bibfnamefont {D.}~\bibnamefont {Alonso}}, \ and\ \bibinfo {author}
		{\bibfnamefont {G.}~\bibnamefont {Adesso}},\ }\href {\doibase
		10.1142/S1230161217400108} {\bibfield  {journal} {\bibinfo  {journal} {Open
				Syst. Inf. Dyn.}\ }\textbf {\bibinfo {volume} {24}},\ \bibinfo {pages}
		{1740010} (\bibinfo {year} {2017})}\BibitemShut {NoStop}%
	\bibitem [{\citenamefont {Asadian}\ \emph {et~al.}(2013)\citenamefont
		{Asadian}, \citenamefont {Manzano}, \citenamefont {Tiersch},\ and\
		\citenamefont {Briegel}}]{asadian2013heat}%
	\BibitemOpen
	\bibfield  {author} {\bibinfo {author} {\bibfnamefont {A.}~\bibnamefont
			{Asadian}}, \bibinfo {author} {\bibfnamefont {D.}~\bibnamefont {Manzano}},
		\bibinfo {author} {\bibfnamefont {M.}~\bibnamefont {Tiersch}}, \ and\
		\bibinfo {author} {\bibfnamefont {H.~J.}\ \bibnamefont {Briegel}},\ }\href
	{\doibase 10.1103/PhysRevE.87.012109} {\bibfield  {journal} {\bibinfo
			{journal} {Phys. Rev. E}\ }\textbf {\bibinfo {volume} {87}},\ \bibinfo
		{pages} {012109} (\bibinfo {year} {2013})}\BibitemShut {NoStop}%
	\bibitem [{\citenamefont {Caldeira}\ and\ \citenamefont
		{Leggett}(1983)}]{caldeira1983path}%
	\BibitemOpen
	\bibfield  {author} {\bibinfo {author} {\bibfnamefont {A.~O.}\ \bibnamefont
			{Caldeira}}\ and\ \bibinfo {author} {\bibfnamefont {A.~J.}\ \bibnamefont
			{Leggett}},\ }\href {\doibase 10.1016/0378-4371(83)90013-4} {\bibfield
		{journal} {\bibinfo  {journal} {Physica A}\ }\textbf {\bibinfo {volume}
			{121}},\ \bibinfo {pages} {587} (\bibinfo {year} {1983})}\BibitemShut
	{NoStop}%
	\bibitem [{\citenamefont {Ford}, \citenamefont {Lewis},\ and\ \citenamefont
		{O'Connell}(1988)}]{ford1988qle}%
	\BibitemOpen
	\bibfield  {author} {\bibinfo {author} {\bibfnamefont {G.~W.}\ \bibnamefont
			{Ford}}, \bibinfo {author} {\bibfnamefont {J.~T.}\ \bibnamefont {Lewis}}, \
		and\ \bibinfo {author} {\bibfnamefont {R.~F.}\ \bibnamefont {O'Connell}},\
	}\href {\doibase 10.1103/PhysRevA.37.4419} {\bibfield  {journal} {\bibinfo
			{journal} {Phys. Rev. A}\ }\textbf {\bibinfo {volume} {37}},\ \bibinfo
		{pages} {4419} (\bibinfo {year} {1988})}\BibitemShut {NoStop}%
	\bibitem [{\citenamefont {Weiss}(2008)}]{weiss2008quantum}%
	\BibitemOpen
	\bibfield  {author} {\bibinfo {author} {\bibfnamefont {U.}~\bibnamefont
			{Weiss}},\ }\href@noop {} {\emph {\bibinfo {title} {{Quantum dissipative
					systems}}}},\ Vol.~\bibinfo {volume} {13}\ (\bibinfo  {publisher} {World
		Scientific Pub Co Inc},\ \bibinfo {year} {2008})\BibitemShut {NoStop}%
	\bibitem [{Note3()}]{Note3}%
	\BibitemOpen
	\bibinfo {note} {Note that in the limit of very large $ \gamma $ this becomes
		$ J_c(\omega ) \sim \protect \frac {\lambda ^2}{\gamma \protect \tmspace
			+\thinmuskip {.1667em}\omega } $.}\BibitemShut {Stop}%
	\bibitem [{\citenamefont {Breuer}\ and\ \citenamefont
		{Petruccione}(2002)}]{Breuer2002}%
	\BibitemOpen
	\bibfield  {author} {\bibinfo {author} {\bibfnamefont {H.}~\bibnamefont
			{Breuer}}\ and\ \bibinfo {author} {\bibfnamefont {F.}~\bibnamefont
			{Petruccione}},\ }\href@noop {} {\emph {\bibinfo {title} {{The Theory of Open
					Quantum Systems}}}}\ (\bibinfo  {publisher} {Oxford University Press, USA},\
	\bibinfo {year} {2002})\BibitemShut {NoStop}%
	\bibitem [{\citenamefont {Suba\ifmmode \mbox{\c{s}}\else \c{s}\fi{}\ifmmode
			\imath \else~\i \fi{}}\ \emph {et~al.}(2012)\citenamefont {Suba\ifmmode
			\mbox{\c{s}}\else \c{s}\fi{}\ifmmode \imath \else~\i \fi{}}, \citenamefont
		{Fleming}, \citenamefont {Taylor},\ and\ \citenamefont
		{Hu}}]{subasi2012equilibrium}%
	\BibitemOpen
	\bibfield  {author} {\bibinfo {author} {\bibfnamefont {Y.}~\bibnamefont
			{Suba\ifmmode \mbox{\c{s}}\else \c{s}\fi{}\ifmmode \imath \else~\i \fi{}}},
		\bibinfo {author} {\bibfnamefont {C.~H.}\ \bibnamefont {Fleming}}, \bibinfo
		{author} {\bibfnamefont {J.~M.}\ \bibnamefont {Taylor}}, \ and\ \bibinfo
		{author} {\bibfnamefont {B.~L.}\ \bibnamefont {Hu}},\ }\href {\doibase
		10.1103/PhysRevE.86.061132} {\bibfield  {journal} {\bibinfo  {journal} {Phys.
				Rev. E}\ }\textbf {\bibinfo {volume} {86}},\ \bibinfo {pages} {061132}
		(\bibinfo {year} {2012})}\BibitemShut {NoStop}%
	\bibitem [{\citenamefont {Gorini}, \citenamefont {Kossakowski},\ and\
		\citenamefont {Sudarshan}(1976)}]{gorini1976completely}%
	\BibitemOpen
	\bibfield  {author} {\bibinfo {author} {\bibfnamefont {V.}~\bibnamefont
			{Gorini}}, \bibinfo {author} {\bibfnamefont {A.}~\bibnamefont {Kossakowski}},
		\ and\ \bibinfo {author} {\bibfnamefont {E.}~\bibnamefont {Sudarshan}},\
	}\href {\doibase 10.1063/1.522979} {\bibfield  {journal} {\bibinfo  {journal}
			{J. Math. Phys.}\ }\textbf {\bibinfo {volume} {17}},\ \bibinfo {pages} {821}
		(\bibinfo {year} {1976})}\BibitemShut {NoStop}%
	\bibitem [{\citenamefont {Lindblad}(1976)}]{lindblad1976generators}%
	\BibitemOpen
	\bibfield  {author} {\bibinfo {author} {\bibfnamefont {G.}~\bibnamefont
			{Lindblad}},\ }\href {\doibase 10.1007/BF01608499} {\bibfield  {journal}
		{\bibinfo  {journal} {Comm. Math. Phys.}\ }\textbf {\bibinfo {volume} {48}},\
		\bibinfo {pages} {119} (\bibinfo {year} {1976})}\BibitemShut {NoStop}%
	\bibitem [{\citenamefont {Joshi}\ \emph {et~al.}(2014)\citenamefont {Joshi},
		\citenamefont {\"Ohberg}, \citenamefont {Cresser},\ and\ \citenamefont
		{Andersson}}]{joshi2014oscillators}%
	\BibitemOpen
	\bibfield  {author} {\bibinfo {author} {\bibfnamefont {C.}~\bibnamefont
			{Joshi}}, \bibinfo {author} {\bibfnamefont {P.}~\bibnamefont {\"Ohberg}},
		\bibinfo {author} {\bibfnamefont {J.~D.}\ \bibnamefont {Cresser}}, \ and\
		\bibinfo {author} {\bibfnamefont {E.}~\bibnamefont {Andersson}},\ }\href
	{\doibase 10.1103/PhysRevA.90.063815} {\bibfield  {journal} {\bibinfo
			{journal} {Phys. Rev. A}\ }\textbf {\bibinfo {volume} {90}},\ \bibinfo
		{pages} {063815} (\bibinfo {year} {2014})}\BibitemShut {NoStop}%
	\bibitem [{\citenamefont {Levy}\ and\ \citenamefont
		{Kosloff}(2014)}]{levy2014local}%
	\BibitemOpen
	\bibfield  {author} {\bibinfo {author} {\bibfnamefont {A.}~\bibnamefont
			{Levy}}\ and\ \bibinfo {author} {\bibfnamefont {R.}~\bibnamefont {Kosloff}},\
	}\href {\doibase 10.1209/0295-5075/107/20004} {\bibfield  {journal} {\bibinfo
			{journal} {Europhys. Lett.}\ }\textbf {\bibinfo {volume} {107}},\ \bibinfo
		{pages} {20004} (\bibinfo {year} {2014})}\BibitemShut {NoStop}%
	\bibitem [{\citenamefont {Stockburger}\ and\ \citenamefont
		{Motz}(2016)}]{stockburger2016thermodynamic}%
	\BibitemOpen
	\bibfield  {author} {\bibinfo {author} {\bibfnamefont {J.~T.}\ \bibnamefont
			{Stockburger}}\ and\ \bibinfo {author} {\bibfnamefont {T.}~\bibnamefont
			{Motz}},\ }\href {\doibase 10.1002/prop.201600067} {\bibfield  {journal}
		{\bibinfo  {journal} {Fortschr. Phys.}\ }\textbf {\bibinfo {volume} {65}},\
		\bibinfo {pages} {6} (\bibinfo {year} {2016})}\BibitemShut {NoStop}%
	\bibitem [{\citenamefont {Ko\l{}ody\'{n}ski}\ \emph {et~al.}(2018)\citenamefont
		{Ko\l{}ody\'{n}ski}, \citenamefont {Brask}, \citenamefont {Perarnau-Llobet},\
		and\ \citenamefont {Bylicka}}]{kolodynski2018adding}%
	\BibitemOpen
	\bibfield  {author} {\bibinfo {author} {\bibfnamefont {J.}~\bibnamefont
			{Ko\l{}ody\'{n}ski}}, \bibinfo {author} {\bibfnamefont {J.~B.}\ \bibnamefont
			{Brask}}, \bibinfo {author} {\bibfnamefont {M.}~\bibnamefont
			{Perarnau-Llobet}}, \ and\ \bibinfo {author} {\bibfnamefont {B.}~\bibnamefont
			{Bylicka}},\ }\href {\doibase 10.1103/PhysRevA.97.062124} {\bibfield
		{journal} {\bibinfo  {journal} {Phys. Rev. A}\ }\textbf {\bibinfo {volume}
			{97}},\ \bibinfo {pages} {062124} (\bibinfo {year} {2018})}\BibitemShut
	{NoStop}%
	\bibitem [{\citenamefont {Wichterich}\ \emph {et~al.}(2007)\citenamefont
		{Wichterich}, \citenamefont {Henrich}, \citenamefont {Breuer}, \citenamefont
		{Gemmer},\ and\ \citenamefont {Michel}}]{PhysRevE.76.031115}%
	\BibitemOpen
	\bibfield  {author} {\bibinfo {author} {\bibfnamefont {H.}~\bibnamefont
			{Wichterich}}, \bibinfo {author} {\bibfnamefont {M.~J.}\ \bibnamefont
			{Henrich}}, \bibinfo {author} {\bibfnamefont {H.-P.}\ \bibnamefont {Breuer}},
		\bibinfo {author} {\bibfnamefont {J.}~\bibnamefont {Gemmer}}, \ and\ \bibinfo
		{author} {\bibfnamefont {M.}~\bibnamefont {Michel}},\ }\href {\doibase
		10.1103/PhysRevE.76.031115} {\bibfield  {journal} {\bibinfo  {journal} {Phys.
				Rev. E}\ }\textbf {\bibinfo {volume} {76}},\ \bibinfo {pages} {031115}
		(\bibinfo {year} {2007})}\BibitemShut {NoStop}%
	\bibitem [{\citenamefont {Trushechkin}\ and\ \citenamefont
		{Volovich}(2016)}]{trushechkin2016perturbative}%
	\BibitemOpen
	\bibfield  {author} {\bibinfo {author} {\bibfnamefont {A.}~\bibnamefont
			{Trushechkin}}\ and\ \bibinfo {author} {\bibfnamefont {I.}~\bibnamefont
			{Volovich}},\ }\href {\doibase 10.1209/0295-5075/113/30005} {\bibfield
		{journal} {\bibinfo  {journal} {Europhys. Lett.}\ }\textbf {\bibinfo {volume}
			{113}},\ \bibinfo {pages} {30005} (\bibinfo {year} {2016})}\BibitemShut
	{NoStop}%
	\bibitem [{\citenamefont {Barra}(2015)}]{barra2015thermodynamic}%
	\BibitemOpen
	\bibfield  {author} {\bibinfo {author} {\bibfnamefont {F.}~\bibnamefont
			{Barra}},\ }\href {\doibase 10.1038/srep14873} {\bibfield  {journal}
		{\bibinfo  {journal} {Sci. Rep.}\ }\textbf {\bibinfo {volume} {5}},\ \bibinfo
		{pages} {14873} (\bibinfo {year} {2015})}\BibitemShut {NoStop}%
	\bibitem [{\citenamefont {Barra}\ and\ \citenamefont
		{Lled\'{o}}(2018)}]{barra2018smallest}%
	\BibitemOpen
	\bibfield  {author} {\bibinfo {author} {\bibfnamefont {F.}~\bibnamefont
			{Barra}}\ and\ \bibinfo {author} {\bibfnamefont {C.}~\bibnamefont
			{Lled\'{o}}},\ }\href {\doibase 10.1140/epjst/e2018-00084-x} {\bibfield
		{journal} {\bibinfo  {journal} {Eur. Phys. J. Spec. Top.}\ }\textbf {\bibinfo
			{volume} {227}},\ \bibinfo {pages} {231} (\bibinfo {year}
		{2018})}\BibitemShut {NoStop}%
	\bibitem [{\citenamefont {De~Chiara}\ \emph {et~al.}(2018)\citenamefont
		{De~Chiara}, \citenamefont {Landi}, \citenamefont {Hewgill}, \citenamefont
		{Reid}, \citenamefont {Ferraro}, \citenamefont {Roncaglia},\ and\
		\citenamefont {Antezza}}]{de2018reconciliation}%
	\BibitemOpen
	\bibfield  {author} {\bibinfo {author} {\bibfnamefont {G.}~\bibnamefont
			{De~Chiara}}, \bibinfo {author} {\bibfnamefont {G.}~\bibnamefont {Landi}},
		\bibinfo {author} {\bibfnamefont {A.}~\bibnamefont {Hewgill}}, \bibinfo
		{author} {\bibfnamefont {B.}~\bibnamefont {Reid}}, \bibinfo {author}
		{\bibfnamefont {A.}~\bibnamefont {Ferraro}}, \bibinfo {author} {\bibfnamefont
			{A.~J.}\ \bibnamefont {Roncaglia}}, \ and\ \bibinfo {author} {\bibfnamefont
			{M.}~\bibnamefont {Antezza}},\ }\href {\doibase 10.1088/1367-2630/aaecee}
	{\bibfield  {journal} {\bibinfo  {journal} {New J. Phys.}\ }\textbf {\bibinfo
			{volume} {20}},\ \bibinfo {pages} {113024} (\bibinfo {year}
		{2018})}\BibitemShut {NoStop}%
	\bibitem [{\citenamefont {Su{\'a}rez}, \citenamefont {Silbey},\ and\
		\citenamefont {Oppenheim}(1992)}]{suarez1992memory}%
	\BibitemOpen
	\bibfield  {author} {\bibinfo {author} {\bibfnamefont {A.}~\bibnamefont
			{Su{\'a}rez}}, \bibinfo {author} {\bibfnamefont {R.}~\bibnamefont {Silbey}},
		\ and\ \bibinfo {author} {\bibfnamefont {I.}~\bibnamefont {Oppenheim}},\
	}\href {\doibase 10.1063/1.463831} {\bibfield  {journal} {\bibinfo  {journal}
			{J. Chem. Phys.}\ }\textbf {\bibinfo {volume} {97}},\ \bibinfo {pages} {5101}
		(\bibinfo {year} {1992})}\BibitemShut {NoStop}%
	\bibitem [{\citenamefont {Gaspard}\ and\ \citenamefont
		{Nagaoka}(1999)}]{gaspard1999slippage}%
	\BibitemOpen
	\bibfield  {author} {\bibinfo {author} {\bibfnamefont {P.}~\bibnamefont
			{Gaspard}}\ and\ \bibinfo {author} {\bibfnamefont {M.}~\bibnamefont
			{Nagaoka}},\ }\href {\doibase 10.1063/1.479867} {\bibfield  {journal}
		{\bibinfo  {journal} {J. Chem. Phys.}\ }\textbf {\bibinfo {volume} {111}},\
		\bibinfo {pages} {5668} (\bibinfo {year} {1999})}\BibitemShut {NoStop}%
	\bibitem [{\citenamefont {Spohn}(1977)}]{spohn1977algebraic}%
	\BibitemOpen
	\bibfield  {author} {\bibinfo {author} {\bibfnamefont {H.}~\bibnamefont
			{Spohn}},\ }\href {\doibase 10.1007/BF00420668} {\bibfield  {journal}
		{\bibinfo  {journal} {Lett. Maths. Phys.}\ }\textbf {\bibinfo {volume} {2}},\
		\bibinfo {pages} {33} (\bibinfo {year} {1977})}\BibitemShut {NoStop}%
	\bibitem [{\citenamefont {Spohn}(1978)}]{spohn1978entropy}%
	\BibitemOpen
	\bibfield  {author} {\bibinfo {author} {\bibfnamefont {H.}~\bibnamefont
			{Spohn}},\ }\href {\doibase 10.1063/1.523789} {\bibfield  {journal} {\bibinfo
			{journal} {J. Math. Phys.}\ }\textbf {\bibinfo {volume} {19}},\ \bibinfo
		{pages} {1227} (\bibinfo {year} {1978})}\BibitemShut {NoStop}%
	\bibitem [{\citenamefont {Ferraro}, \citenamefont {Olivares},\ and\
		\citenamefont {Paris}(2005)}]{0503237v1}%
	\BibitemOpen
	\bibfield  {author} {\bibinfo {author} {\bibfnamefont {A.}~\bibnamefont
			{Ferraro}}, \bibinfo {author} {\bibfnamefont {S.}~\bibnamefont {Olivares}}, \
		and\ \bibinfo {author} {\bibfnamefont {M.}~\bibnamefont {Paris}},\ }\href
	{https://arxiv.org/abs/quant-ph/0503237} {\emph {\bibinfo {title} {{Gaussian
					states in continuous variable quantum information}}}},\ edited by\ \bibinfo
	{editor} {\bibfnamefont {I.}~\bibnamefont {88-7088-483-X}}\ (\bibinfo
	{publisher} {Bibliopolis, Napoli},\ \bibinfo {year} {2005})\BibitemShut
	{NoStop}%
	\bibitem [{\citenamefont {Alicki}(1979)}]{alicki1979engine}%
	\BibitemOpen
	\bibfield  {author} {\bibinfo {author} {\bibfnamefont {R.}~\bibnamefont
			{Alicki}},\ }\href {\doibase 10.1088/0305-4470/12/5/007} {\bibfield
		{journal} {\bibinfo  {journal} {J. Phys. A}\ }\textbf {\bibinfo {volume}
			{12}},\ \bibinfo {pages} {L103} (\bibinfo {year} {1979})}\BibitemShut
	{NoStop}%
	\bibitem [{\citenamefont {Kosloff}\ and\ \citenamefont
		{Levy}(2014)}]{Kosloff2014}%
	\BibitemOpen
	\bibfield  {author} {\bibinfo {author} {\bibfnamefont {R.}~\bibnamefont
			{Kosloff}}\ and\ \bibinfo {author} {\bibfnamefont {A.}~\bibnamefont {Levy}},\
	}\href {\doibase 10.1146/annurev-physchem-040513-103724} {\bibfield
		{journal} {\bibinfo  {journal} {Anual Rev. Phys. Chem.}\ }\textbf {\bibinfo
			{volume} {65}},\ \bibinfo {pages} {365} (\bibinfo {year} {2014})}\BibitemShut
	{NoStop}%
	\bibitem [{Note4()}]{Note4}%
	\BibitemOpen
	\bibinfo {note} {Indeed, all the standard quantum-thermodynamic arguments
		based on the contractivity of the dissipative dynamics and uniqueness of the
		(thermal) fixed point of any of the dissipators of a GKLS equation \cite
		{spohn1978entropy,alicki1979engine} hold regardless of whether or not the
		Lamb shift is included in $ \protect \mathcal {L}_i $. In particular, one
		\protect \textit {always} finds $ \protect \mathcal {L}_i\protect \tmspace
		+\thinmuskip {.1667em}e^{-\protect \pmb H_N/T_i} = 0 $, i.e., the fixed point
		of the dynamics is a thermal state with respect to the \protect \textit
		{unshifted} Hamiltonian.}\BibitemShut {Stop}%
	\bibitem [{\citenamefont {Riseborough}, \citenamefont {Hanggi},\ and\
		\citenamefont {Weiss}(1985)}]{riseborough1985exact}%
	\BibitemOpen
	\bibfield  {author} {\bibinfo {author} {\bibfnamefont {P.}~\bibnamefont
			{Riseborough}}, \bibinfo {author} {\bibfnamefont {P.}~\bibnamefont {Hanggi}},
		\ and\ \bibinfo {author} {\bibfnamefont {U.}~\bibnamefont {Weiss}},\ }\href
	{\doibase 10.1103/PhysRevA.31.471} {\bibfield  {journal} {\bibinfo  {journal}
			{Phys. Rev. A}\ }\textbf {\bibinfo {volume} {31}},\ \bibinfo {pages} {471}
		(\bibinfo {year} {1985})}\BibitemShut {NoStop}%
	\bibitem [{\citenamefont {Ludwig}, \citenamefont {Hammerer},\ and\
		\citenamefont {Marquardt}(2010)}]{ludwig2010entanglement}%
	\BibitemOpen
	\bibfield  {author} {\bibinfo {author} {\bibfnamefont {M.}~\bibnamefont
			{Ludwig}}, \bibinfo {author} {\bibfnamefont {K.}~\bibnamefont {Hammerer}}, \
		and\ \bibinfo {author} {\bibfnamefont {F.}~\bibnamefont {Marquardt}},\ }\href
	{\doibase 10.1103/PhysRevA.82.012333} {\bibfield  {journal} {\bibinfo
			{journal} {Phys. Rev. A}\ }\textbf {\bibinfo {volume} {82}},\ \bibinfo
		{pages} {012333} (\bibinfo {year} {2010})}\BibitemShut {NoStop}%
	\bibitem [{\citenamefont {Fleming}, \citenamefont {Roura},\ and\ \citenamefont
		{Hu}(2011)}]{fleming2011exact}%
	\BibitemOpen
	\bibfield  {author} {\bibinfo {author} {\bibfnamefont {C.}~\bibnamefont
			{Fleming}}, \bibinfo {author} {\bibfnamefont {A.}~\bibnamefont {Roura}}, \
		and\ \bibinfo {author} {\bibfnamefont {B.}~\bibnamefont {Hu}},\ }\href
	{\doibase 10.1016/j.aop.2010.12.003} {\bibfield  {journal} {\bibinfo
			{journal} {Ann. Phys. (N. Y.)}\ }\textbf {\bibinfo {volume} {326}},\ \bibinfo
		{pages} {1207} (\bibinfo {year} {2011})}\BibitemShut {NoStop}%
	\bibitem [{\citenamefont {Correa}, \citenamefont {Valido},\ and\ \citenamefont
		{Alonso}(2012)}]{correa2012asymptotic}%
	\BibitemOpen
	\bibfield  {author} {\bibinfo {author} {\bibfnamefont {L.~A.}\ \bibnamefont
			{Correa}}, \bibinfo {author} {\bibfnamefont {A.~A.}\ \bibnamefont {Valido}},
		\ and\ \bibinfo {author} {\bibfnamefont {D.}~\bibnamefont {Alonso}},\ }\href
	{\doibase 10.1103/PhysRevA.86.012110} {\bibfield  {journal} {\bibinfo
			{journal} {Phys. Rev. A}\ }\textbf {\bibinfo {volume} {86}},\ \bibinfo
		{pages} {012110} (\bibinfo {year} {2012})}\BibitemShut {NoStop}%
	\bibitem [{\citenamefont {Martinez}\ and\ \citenamefont
		{Paz}(2013)}]{martinez2013dynamics}%
	\BibitemOpen
	\bibfield  {author} {\bibinfo {author} {\bibfnamefont {E.~A.}\ \bibnamefont
			{Martinez}}\ and\ \bibinfo {author} {\bibfnamefont {J.~P.}\ \bibnamefont
			{Paz}},\ }\href {\doibase 10.1103/PhysRevLett.110.130406} {\bibfield
		{journal} {\bibinfo  {journal} {Phys. Rev. Lett.}\ }\textbf {\bibinfo
			{volume} {110}},\ \bibinfo {pages} {130406} (\bibinfo {year}
		{2013})}\BibitemShut {NoStop}%
	\bibitem [{\citenamefont {Valido}, \citenamefont {Ruiz},\ and\ \citenamefont
		{Alonso}(2015)}]{valido2015currents}%
	\BibitemOpen
	\bibfield  {author} {\bibinfo {author} {\bibfnamefont {A.~A.}\ \bibnamefont
			{Valido}}, \bibinfo {author} {\bibfnamefont {A.}~\bibnamefont {Ruiz}}, \ and\
		\bibinfo {author} {\bibfnamefont {D.}~\bibnamefont {Alonso}},\ }\href
	{\doibase 10.1103/PhysRevE.91.062123} {\bibfield  {journal} {\bibinfo
			{journal} {Phys. Rev. E}\ }\textbf {\bibinfo {volume} {91}},\ \bibinfo
		{pages} {062123} (\bibinfo {year} {2015})}\BibitemShut {NoStop}%
	\bibitem [{\citenamefont {Freitas}\ and\ \citenamefont
		{Paz}(2014)}]{freitas2014analytic}%
	\BibitemOpen
	\bibfield  {author} {\bibinfo {author} {\bibfnamefont {N.}~\bibnamefont
			{Freitas}}\ and\ \bibinfo {author} {\bibfnamefont {J.~P.}\ \bibnamefont
			{Paz}},\ }\href {\doibase 10.1103/PhysRevE.90.042128} {\bibfield  {journal}
		{\bibinfo  {journal} {Phys. Rev. E}\ }\textbf {\bibinfo {volume} {90}},\
		\bibinfo {pages} {042128} (\bibinfo {year} {2014})}\BibitemShut {NoStop}%
	\bibitem [{\citenamefont {Ford}, \citenamefont {Kac},\ and\ \citenamefont
		{Mazur}(1965)}]{ford1965statistical}%
	\BibitemOpen
	\bibfield  {author} {\bibinfo {author} {\bibfnamefont {G.}~\bibnamefont
			{Ford}}, \bibinfo {author} {\bibfnamefont {M.}~\bibnamefont {Kac}}, \ and\
		\bibinfo {author} {\bibfnamefont {P.}~\bibnamefont {Mazur}},\ }\href
	{\doibase 10.1063/1.1704304} {\bibfield  {journal} {\bibinfo  {journal} {J.
				Math. Phys.}\ }\textbf {\bibinfo {volume} {6}},\ \bibinfo {pages} {504}
		(\bibinfo {year} {1965})}\BibitemShut {NoStop}%
	\bibitem [{\citenamefont {Correa}\ \emph {et~al.}(2017)\citenamefont {Correa},
		\citenamefont {Perarnau-Llobet}, \citenamefont {Hovhannisyan}, \citenamefont
		{Hern{\'a}ndez-Santana}, \citenamefont {Mehboudi},\ and\ \citenamefont
		{Sanpera}}]{correa2017strong}%
	\BibitemOpen
	\bibfield  {author} {\bibinfo {author} {\bibfnamefont {L.~A.}\ \bibnamefont
			{Correa}}, \bibinfo {author} {\bibfnamefont {M.}~\bibnamefont
			{Perarnau-Llobet}}, \bibinfo {author} {\bibfnamefont {K.~V.}\ \bibnamefont
			{Hovhannisyan}}, \bibinfo {author} {\bibfnamefont {S.}~\bibnamefont
			{Hern{\'a}ndez-Santana}}, \bibinfo {author} {\bibfnamefont {M.}~\bibnamefont
			{Mehboudi}}, \ and\ \bibinfo {author} {\bibfnamefont {A.}~\bibnamefont
			{Sanpera}},\ }\href {\doibase 10.1103/PhysRevA.96.062103} {\bibfield
		{journal} {\bibinfo  {journal} {Phys. Rev. A}\ }\textbf {\bibinfo {volume}
			{96}},\ \bibinfo {pages} {062103} (\bibinfo {year} {2017})}\BibitemShut
	{NoStop}%
	\bibitem [{\citenamefont {Valido}, \citenamefont {Correa},\ and\ \citenamefont
		{Alonso}(2013)}]{PhysRevA.88.012309}%
	\BibitemOpen
	\bibfield  {author} {\bibinfo {author} {\bibfnamefont {A.~A.}\ \bibnamefont
			{Valido}}, \bibinfo {author} {\bibfnamefont {L.~A.}\ \bibnamefont {Correa}},
		\ and\ \bibinfo {author} {\bibfnamefont {D.}~\bibnamefont {Alonso}},\ }\href
	{\doibase 10.1103/PhysRevA.88.012309} {\bibfield  {journal} {\bibinfo
			{journal} {Phys. Rev. A}\ }\textbf {\bibinfo {volume} {88}},\ \bibinfo
		{pages} {012309} (\bibinfo {year} {2013})}\BibitemShut {NoStop}%
	\bibitem [{\citenamefont {Freitas}\ and\ \citenamefont
		{Paz}(2017)}]{freitas2017fundamental}%
	\BibitemOpen
	\bibfield  {author} {\bibinfo {author} {\bibfnamefont {N.}~\bibnamefont
			{Freitas}}\ and\ \bibinfo {author} {\bibfnamefont {J.~P.}\ \bibnamefont
			{Paz}},\ }\href {\doibase 10.1103/PhysRevE.95.012146} {\bibfield  {journal}
		{\bibinfo  {journal} {Phys. Rev. E}\ }\textbf {\bibinfo {volume} {95}},\
		\bibinfo {pages} {012146} (\bibinfo {year} {2017})}\BibitemShut {NoStop}%
	\bibitem [{\citenamefont {Motz}, \citenamefont {Ankerhold},\ and\ \citenamefont
		{Stockburger}(2017)}]{motz2017currents}%
	\BibitemOpen
	\bibfield  {author} {\bibinfo {author} {\bibfnamefont {T.}~\bibnamefont
			{Motz}}, \bibinfo {author} {\bibfnamefont {J.}~\bibnamefont {Ankerhold}}, \
		and\ \bibinfo {author} {\bibfnamefont {J.~T.}\ \bibnamefont {Stockburger}},\
	}\href {https://doi.org/10.1088/1367-2630/aa68bd} {\bibfield  {journal}
		{\bibinfo  {journal} {New J. Phys.}\ }\textbf {\bibinfo {volume} {19}},\
		\bibinfo {pages} {053013} (\bibinfo {year} {2017})}\BibitemShut {NoStop}%
	\bibitem [{\citenamefont {Motz}\ \emph {et~al.}(2018)\citenamefont {Motz},
		\citenamefont {Wiedmann}, \citenamefont {Stockburger},\ and\ \citenamefont
		{Ankerhold}}]{motz2018rectification}%
	\BibitemOpen
	\bibfield  {author} {\bibinfo {author} {\bibfnamefont {T.}~\bibnamefont
			{Motz}}, \bibinfo {author} {\bibfnamefont {M.}~\bibnamefont {Wiedmann}},
		\bibinfo {author} {\bibfnamefont {J.~T.}\ \bibnamefont {Stockburger}}, \ and\
		\bibinfo {author} {\bibfnamefont {J.}~\bibnamefont {Ankerhold}},\ }\href
	{https://doi.org/10.1088/1367-2630/aaea90} {\bibfield  {journal} {\bibinfo
			{journal} {New J. Phys.}\ }\textbf {\bibinfo {volume} {20}},\ \bibinfo
		{pages} {113020} (\bibinfo {year} {2018})}\BibitemShut {NoStop}%
	\bibitem [{\citenamefont {Uhlmann}(1976)}]{uhlmann1976transition}%
	\BibitemOpen
	\bibfield  {author} {\bibinfo {author} {\bibfnamefont {A.}~\bibnamefont
			{Uhlmann}},\ }\href {\doibase 10.1016/0034-4877(76)90060-4} {\bibfield
		{journal} {\bibinfo  {journal} {Rep. Math. Phys.}\ }\textbf {\bibinfo
			{volume} {9}},\ \bibinfo {pages} {273} (\bibinfo {year} {1976})}\BibitemShut
	{NoStop}%
	\bibitem [{\citenamefont {Banchi}, \citenamefont {Braunstein},\ and\
		\citenamefont {Pirandola}(2015)}]{banchi2005fidelity}%
	\BibitemOpen
	\bibfield  {author} {\bibinfo {author} {\bibfnamefont {L.}~\bibnamefont
			{Banchi}}, \bibinfo {author} {\bibfnamefont {S.~L.}\ \bibnamefont
			{Braunstein}}, \ and\ \bibinfo {author} {\bibfnamefont {S.}~\bibnamefont
			{Pirandola}},\ }\href {\doibase 10.1103/PhysRevLett.115.260501} {\bibfield
		{journal} {\bibinfo  {journal} {Phys. Rev. Lett.}\ }\textbf {\bibinfo
			{volume} {115}},\ \bibinfo {pages} {260501} (\bibinfo {year}
		{2015})}\BibitemShut {NoStop}%
	\bibitem [{\citenamefont {Ferraro}, \citenamefont {Garc{\'\i}a-Saez},\ and\
		\citenamefont {Ac{\'\i}n}(2012)}]{ferraro2012intensive}%
	\BibitemOpen
	\bibfield  {author} {\bibinfo {author} {\bibfnamefont {A.}~\bibnamefont
			{Ferraro}}, \bibinfo {author} {\bibfnamefont {A.}~\bibnamefont
			{Garc{\'\i}a-Saez}}, \ and\ \bibinfo {author} {\bibfnamefont
			{A.}~\bibnamefont {Ac{\'\i}n}},\ }\href
	{https://doi.org/10.1209/0295-5075/98/10009
		https://doi.org/10.1209/0295-5075/98/10009} {\bibfield  {journal} {\bibinfo
			{journal} {EPL (Europhysics Letters)}\ }\textbf {\bibinfo {volume} {98}},\
		\bibinfo {pages} {10009} (\bibinfo {year} {2012})}\BibitemShut {NoStop}%
	\bibitem [{\citenamefont {Garc\'{\i}a-Saez}, \citenamefont {Ferraro},\ and\
		\citenamefont {Ac\'{\i}n}(2009)}]{PhysRevA.79.052340}%
	\BibitemOpen
	\bibfield  {author} {\bibinfo {author} {\bibfnamefont {A.}~\bibnamefont
			{Garc\'{\i}a-Saez}}, \bibinfo {author} {\bibfnamefont {A.}~\bibnamefont
			{Ferraro}}, \ and\ \bibinfo {author} {\bibfnamefont {A.}~\bibnamefont
			{Ac\'{\i}n}},\ }\href {\doibase 10.1103/PhysRevA.79.052340} {\bibfield
		{journal} {\bibinfo  {journal} {Phys. Rev. A}\ }\textbf {\bibinfo {volume}
			{79}},\ \bibinfo {pages} {052340} (\bibinfo {year} {2009})}\BibitemShut
	{NoStop}%
	\bibitem [{\citenamefont {Kliesch}\ \emph {et~al.}(2014)\citenamefont
		{Kliesch}, \citenamefont {Gogolin}, \citenamefont {Kastoryano}, \citenamefont
		{Riera},\ and\ \citenamefont {Eisert}}]{PhysRevX.4.031019}%
	\BibitemOpen
	\bibfield  {author} {\bibinfo {author} {\bibfnamefont {M.}~\bibnamefont
			{Kliesch}}, \bibinfo {author} {\bibfnamefont {C.}~\bibnamefont {Gogolin}},
		\bibinfo {author} {\bibfnamefont {M.~J.}\ \bibnamefont {Kastoryano}},
		\bibinfo {author} {\bibfnamefont {A.}~\bibnamefont {Riera}}, \ and\ \bibinfo
		{author} {\bibfnamefont {J.}~\bibnamefont {Eisert}},\ }\href {\doibase
		10.1103/PhysRevX.4.031019} {\bibfield  {journal} {\bibinfo  {journal} {Phys.
				Rev. X}\ }\textbf {\bibinfo {volume} {4}},\ \bibinfo {pages} {031019}
		(\bibinfo {year} {2014})}\BibitemShut {NoStop}%
	\bibitem [{\citenamefont {Hern{\'a}ndez-Santana}\ \emph
		{et~al.}(2015)\citenamefont {Hern{\'a}ndez-Santana}, \citenamefont {Riera},
		\citenamefont {Hovhannisyan}, \citenamefont {Perarnau-Llobet}, \citenamefont
		{Tagliacozzo},\ and\ \citenamefont {Ac{\'\i}n}}]{hernandez2015locality}%
	\BibitemOpen
	\bibfield  {author} {\bibinfo {author} {\bibfnamefont {S.}~\bibnamefont
			{Hern{\'a}ndez-Santana}}, \bibinfo {author} {\bibfnamefont {A.}~\bibnamefont
			{Riera}}, \bibinfo {author} {\bibfnamefont {K.~V.}\ \bibnamefont
			{Hovhannisyan}}, \bibinfo {author} {\bibfnamefont {M.}~\bibnamefont
			{Perarnau-Llobet}}, \bibinfo {author} {\bibfnamefont {L.}~\bibnamefont
			{Tagliacozzo}}, \ and\ \bibinfo {author} {\bibfnamefont {A.}~\bibnamefont
			{Ac{\'\i}n}},\ }\href {https://doi.org/10.1088/1367-2630/17/8/085007}
	{\bibfield  {journal} {\bibinfo  {journal} {New J. Phys.}\ }\textbf {\bibinfo
			{volume} {17}},\ \bibinfo {pages} {085007} (\bibinfo {year}
		{2015})}\BibitemShut {NoStop}%
	\bibitem [{\citenamefont {Strasberg}\ and\ \citenamefont
		{Esposito}(2017)}]{strasberg2017stochastic}%
	\BibitemOpen
	\bibfield  {author} {\bibinfo {author} {\bibfnamefont {P.}~\bibnamefont
			{Strasberg}}\ and\ \bibinfo {author} {\bibfnamefont {M.}~\bibnamefont
			{Esposito}},\ }\href {\doibase 10.1103/PhysRevE.95.062101} {\bibfield
		{journal} {\bibinfo  {journal} {Phys. Rev. E}\ }\textbf {\bibinfo {volume}
			{95}},\ \bibinfo {pages} {062101} (\bibinfo {year} {2017})}\BibitemShut
	{NoStop}%
	\bibitem [{\citenamefont {Newman}, \citenamefont {Mintert},\ and\ \citenamefont
		{Nazir}(2017)}]{newman2017performance}%
	\BibitemOpen
	\bibfield  {author} {\bibinfo {author} {\bibfnamefont {D.}~\bibnamefont
			{Newman}}, \bibinfo {author} {\bibfnamefont {F.}~\bibnamefont {Mintert}}, \
		and\ \bibinfo {author} {\bibfnamefont {A.}~\bibnamefont {Nazir}},\ }\href
	{\doibase 10.1103/PhysRevE.95.032139} {\bibfield  {journal} {\bibinfo
			{journal} {Phys. Rev. E}\ }\textbf {\bibinfo {volume} {95}},\ \bibinfo
		{pages} {032139} (\bibinfo {year} {2017})}\BibitemShut {NoStop}%
	\bibitem [{\citenamefont {Tamascelli}\ \emph {et~al.}(2018)\citenamefont
		{Tamascelli}, \citenamefont {Smirne}, \citenamefont {Huelga},\ and\
		\citenamefont {Plenio}}]{PhysRevLett.120.030402}%
	\BibitemOpen
	\bibfield  {author} {\bibinfo {author} {\bibfnamefont {D.}~\bibnamefont
			{Tamascelli}}, \bibinfo {author} {\bibfnamefont {A.}~\bibnamefont {Smirne}},
		\bibinfo {author} {\bibfnamefont {S.~F.}\ \bibnamefont {Huelga}}, \ and\
		\bibinfo {author} {\bibfnamefont {M.~B.}\ \bibnamefont {Plenio}},\ }\href
	{\doibase 10.1103/PhysRevLett.120.030402} {\bibfield  {journal} {\bibinfo
			{journal} {Phys. Rev. Lett.}\ }\textbf {\bibinfo {volume} {120}},\ \bibinfo
		{pages} {030402} (\bibinfo {year} {2018})}\BibitemShut {NoStop}%
\end{thebibliography}

\end{document}